\documentclass{article}

\usepackage{arxiv}

\usepackage[utf8]{inputenc} 
\usepackage[T1]{fontenc}    
\usepackage{hyperref}       
\usepackage{url}            
\usepackage{booktabs}       
\usepackage{amsfonts}       
\usepackage{nicefrac}       
\usepackage{microtype}      
\usepackage{lipsum}
\usepackage{graphicx}
\graphicspath{ {./images/} }

\usepackage{amssymb}
\usepackage{amsthm}
\usepackage{mathrsfs}
\usepackage{amsmath}
\usepackage{float}
\usepackage[linewidth=1pt]{mdframed}
\usepackage{tabularx}
\DeclareSymbolFont{bbold}{U}{bbold}{m}{n}
\DeclareSymbolFontAlphabet{\mathbbold}{bbold}
\usepackage{subcaption}

\def\bfe{{\bf e}}

\def\bfn{{\bf n}}

\def\bfr{{\bf r}}

\def\bfv{{\bf v}}

\def\bfx{{\bf x}}

\def\bfA{{\bf A}}
\def\bfE{{\bf E}}

\def\bfC{{\bf C}}
\def\bfD{{\bf D}}

\def\bfL{{\bf L}}
\def\bfM{{\bf M}}
\def\bfN{{\bf N}}
\def\bfP{{\bf P}}

\def\bfR{{\bf R}}
\def\bfS{{\bf S}}

\def\bfX{{\bf X}}
\def\bfF{{\bf F}}

\def\bfV{{\bf V}}
\def\bfW{{\bf W}}

\def\eps{\epsilon}

\def\l{\lambda}

\def\sig{\sigma}
\def\Sig{\Sigma}

\def\bfzero{\mbox{\boldmath $0$}}

\def\bfsig{\mbox{\boldmath $\sigma$}}
\def\bfSig{\mbox{\boldmath $\Sigma$}}

\def\bfdel{\mbox{\boldmath $\delta$}}

\def\bftau{\mbox{\boldmath $\tau$}}

\def\e0{\varepsilon_0}

\def\oF{\overline{F}}
\def\oG{\overline{G}}
\def\oI{\overline{I}}
\def\oP{\overline{P}}

\def\osig{\overline{\sig}}
\def\oSig{\overline{\Sig}}

\def\obfF{\overline{\bfF}}

\def\obfP{\overline{\bfP}}

\def\ol{\overline{\lambda}}
\def\ogamma{\overline{\gamma}}

\def\obfsig{\overline{\bfsig}}
\def\obfSig{\overline{\bfSig}}
\def\obftau{\overline{\bftau}}
\def\p{\partial}

\def\oJ{\overline{J}}

\def\s0{\sigma_0}

\def\bfK{\mbox{\boldmath $K$}}

\def\bk0{\mathbb{0}}

\def\oeps{\overline{\varepsilon}}

\def\p{\partial}

\def\bftauj{{\mathop {\bftau}\limits^\triangledown}}
\def\bftaut{{\mathop {\bftau}\limits^\triangle}}

\def\bmA{\pmb{\mathcal{A}}}

\def\bsC{\pmb{\mathscr{C}}}

\def\bsL{\pmb{\mathscr{L}}}
\def\scT{{\mathscr{T}}}
\def\bsT{\pmb{\mathscr{T}}}

\def\bmL{\pmb{\mathcal{L}}}
\def\bmH{\pmb{\mathcal{H}}}

\def\bmI{\pmb{\mathcal{I}}}

\title{The effect of fiber plasticity on domain formation in soft biological composites---Part I: A bifurcation analysis}

\author{
 Michalis Agoras \\
  Department of Mechanical Engineering\\
  University of Thessaly, Volos, Greece \\
  \texttt{agoras@uth.gr} \\
   \And
 Fernanda F. Fontenele \\
  Sibley School of Mechanical and Aerospace Engineering\\
  Cornell University, Ithaca, NY, USA \\ 
  \And
 Nikolaos Bouklas \\
  Sibley School of Mechanical and Aerospace Engineering\\
  Cornell University, Ithaca, NY, USA \\ 
  \& Pasteur Labs, Brooklyn, NY\\
  \texttt{nbouklas@cornell.edu} \\
}

\begin{document}
\maketitle\footnote{MA and FFF have an equal contribution to the manuscript.}
\begin{abstract}

The main objective of this work is to shed light on the effect of fiber plasticity on the macroscopic response and domain formation in soft biological composites. 
This goal is pursued by analyzing the plane-strain response of two-phase laminates.
In the context of this problem, the effect of fiber plasticity is accounted for by allowing the elastically stiffer layers (``fiber'' phase) to also yield plastically and by taking the soft layers (``matrix'' phase) to be purely elastic solids.
The analysis is carried out at finite elastic and plastic strains, but it is restricted to unidirectional, non-monotonic loading paths, applied by initially increasing the macroscopic stretch along the direction of the layers up to a prescribed maximum value and then decreasing the same stretch down to a minimum value. 
A simple expression is derived for the critical conditions at which the homogenized behavior of the laminate loses strong ellipticity for the first time along the loading path. 
The relevance of this result stems from the fact that the loss of macroscopic ellipticity of these composites is known to coincide with the onset of bifurcations of the long-wavelength type.
It follows from this result that, just like hyperelastic laminates, elastoplastic laminates may lose macroscopic ellipticity whenever their incremental strength in shear perpendicular to the layers vanishes for the first time. 
For situations in which loss of macroscopic ellipticity does take place, a corresponding post-bifurcation solution for the homogenized behavior of the laminate is computed.
The deformed state of the material described by this solution is characterized by twin lamellar domains that are formed at a length scale much larger than the width of the original, microscopic layers, but still much smaller than the overall dimensions of the macroscopic specimen under consideration.

\end{abstract}


\section{Introduction}\label{Intro}

\begin{figure}[h]
\begin{center} 
    \includegraphics[width=15 cm]{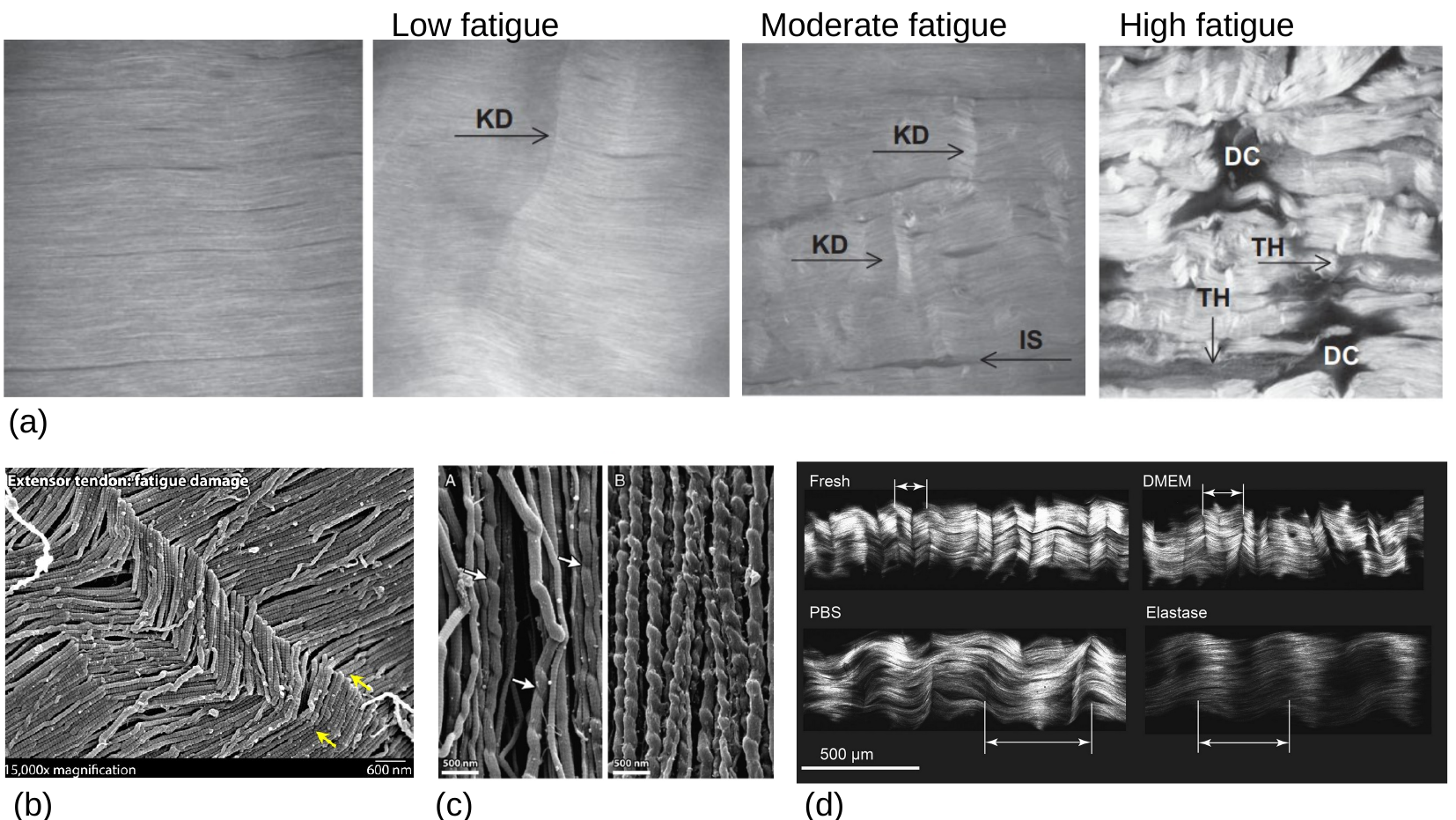}
  \end{center}

\caption {\footnotesize
{(a) Multiphoton microscopy of rat patellar tendon under cyclic tensile loading. Control specimen is compared to three stages of fatigue induced damage progression, indicated as low, moderate and high damage (adapted from \cite{fung2010early}); $\mathrm{FOV}=400\mathrm{mm}$. (b) Extensor tendons from bovine forelimb showed macroscopic kink-bands, while individual fibrils had signs of discrete plasticity (kinks at fibril level) after tensile cyclic loading (adapted from  \cite{herod2016collagen}). (c) Tensile overload causes discrete plasticity in collagen fibrils of bovine tail tendon. (Left)  Overloading tendon causes fibrils to develop permanent micro-necks (arrows) that repeat every $\approx 300-800\mathrm{nm}$ along
their length; (right) cyclic loading intensifies this phenomenon (adapted from  \cite{veres2014mechanically}). (d)  Comparing morphology of control tendon specimens to one that has undergone elastase treatment (adapted from \cite{grant2015mechanical}).}}
  \label{Tendon}

\end{figure}

%
%
Dense connective tissues such as tendons and ligaments may be idealized as unidirectional, fiber-reinforced composites, consisting of stiff and parallel fibers embedded in a soft matrix.  
The matrix phase in these materials is hydrated and it is composed mainly of elastin, small leucin-rich repeat proteoglycans, biclycan and decorin, glycosaminoglycans, and tenocytes (tendon cells), whereas the fibers are hierarchically structured from tropocollagen molecules to fibrils \cite{reese2015tendons}.
Under natural operating conditions,
these materials are typically subjected to tensile oading/unloading cycles along the fiber direction and often fail to recover their original shape after unloading. 
%
Certain permanent, chevron-like deformation patterns that are observed in tendons have been associated in the literature \cite{fung2010early,andarawis2011tendon} with the development of diseases, such as tendinopathy, and mainly treated as an early marker of disease and ultimate failure.
These patterns, as illustrated  in Fig. \ref{Tendon}(a-b,d), are characterized by two distinct types of layer-like ``domains,'' which are apparently formed through collective fiber kinking over surfaces that are oriented (on average) perpendicular to the undeformed direction of the fibers.
The fact that these patterns are unrecoverable after unloading suggests the presence of inelastic deformation mechanisms in the material. 
Indeed, it is known from further experimental evidence \cite{baldwin2016characterization,tang2010deformation,veres2013repeated,fontenele2023understanding} that collagen fibers may undergo significant plastic deformations during macroscopic loading / unloading of tendons.
Further evidence of plasticity in collagen fibers, manifesting at the fibril level is also observed as seen in Fig. \ref{Tendon}(c), where periodic micro-necking occurs during fatigue loading. 
Understanding the origin and evolution of these complex deformation patterns is a challenging and important problem from both the theoretical and practical point of view, which remains largely unexplored to this day. 
%

%
The main objective of this work is to shed light on the effect of fiber plasticity on the macroscopic response and domain formation in soft and biological composites. 
To this end, we treat domain formation in these materials as a bifurcation phenomenon and make use of exact results for periodic elastoplastic laminates, combined with additional well-established results from the literature, which are briefly reviewed next, to study the onset and evolution of these domains. 
It should be remarked that because of their analytical tractability, layered composites are commonly used in the literature as two-dimensional approximations of actual fiber-reinforced composites. 

Substantial progress has been made over the past several decades towards understanding the compressive failure of unidirectional, fiber-reinforced composites that are used in technological applications, including glass- and carbon-reinforced polymers.
The failure of these materials under uniaxial compression along the fiber direction is known to occur through strain localization in narrow kink bands.
The first important contribution to this problem is due to \cite{R65},  who identified fiber micro-buckling as the onset of failure and derived a simple formula for the associated critical stress, assuming that the matrix phase is purely elastic.
Subsequent developments by \cite{A72} and \cite{B83} highlighted the crucial role of fiber misalignment (with respect to the loading axis) in triggering plastic deformations in the matrix and in enhancing significantly the propagation of the kink bands. 
A more detailed analysis of various aspects of the plastic kinking process, including the effects of inclined kink bands and combined axial compression and shear stresses, can be found in the work by \cite{BF93}.
Building on these pioneering developments, an extensive amount of research has been dedicated towards a deeper understanding of the compressive failure of fiber-reinforced composites, a detailed review of which lies beyond the scope of the present article.
However, in connection with the imperfection analysis presented in a forthcoming article (Part II), it is relevant to make a reference here to the work by \cite{KAPL95}. These authors studied experimentally the compressive failure of a graphite-epoxy composite and carried out full-field numerical simulations of the experimental process by modeling the actual fiber-reinforced composite as a two-phase laminate, comprised of elastoplastic (matrix) and purely elastic layers (fibers) with a small initial waviness. 
The latter geometric imperfection has been introduced in the model in  order to trigger the formation of kink-bands.
Corresponding three dimensional calculations for periodic composites comprised of hexagonal arrays of cylindrical fibers with circular cross sections have been performed by \cite{HVK98}.
Along the same lines, \cite{VHK00} studied the effect of combined axial compression and shear on the development of kink-band instabilities in these composites. 
It is important to emphasize that the fibers in these technological composites are brittle, as opposed to their counterparts in soft biological composites, which are ductile. It should also be remarked, however, that in both cases the fiber phase is elastically stiffer than the matrix.
The bifurcation problem for the incremental behavior of periodic laminates with nonlinear constituents under plane-strain loading paths has been first investigated by \cite{TM85}. 
An important finding of this work is that the onset of a bifurcation to a long-wavelength (compared with the unit-cell size) mode coincides with the loss of ellipticity (LOE) of the incremental equations of equilibrium for the homogenized behavior of the laminate.
\cite{GMT93} have shown that this result holds for all composite materials with periodic microstructures and hyperelastic constituents.
For the special case of two-phase laminates subjected to plane-strain unidirectional loading along the layer direction, which is of particular interest in this work, it has been found (\cite{TM85}, \cite{GMT93}, \cite{NT04}) that the critical bifurcation mode, i.e., the first bifurcation that occurs as the material is monotonically loaded from the unstressed state, is always of the long-wavelength type, provided that the concentration of the stiff phase is not too small (typically, when it is greater than $10\%$).
Motivated by these findings, the LOE condition has been used in numerous investigations as a criterion for the onset of macroscopic instabilities in hyperelastic fiber-reinforced composites, including the works by \cite{QP97} and \cite{MO03}, who studied the LOE of phenomenological models for these materials, as well as the work by \cite{ALPPC09b}, who made use of the corresponding homogenization models of \cite{DeBHS06} and \cite{ALPPC09a}. 
The role of microstructure evolution on the macroscopic LOE of two-phase Neo-Hookean laminates has been studied by \cite{LPPC2009}. 
More recently, \cite{SdATW16} determined the post-bifurcation behavior of hyperelastic laminates under aligned, plane-strain compression and examined the effect of varying the local material properties of the layers on the overall stability of the composite. 
They found that the macroscopic behavior of the laminate is always stable for Neo-Hookean constituents and it becomes unstable only in situations where the incremental strength of the soft phase exhibits a significant deterioration with increasing strain, a behavior which is reminiscent of an elastoplastic rather than a hyperelastic solid.
A general framework for computing the homogenized, long-wavelength, post-bifurcation behavior of hyperelastic composites in terms of the quasiconvexification or ``relaxation'' of the principal solution for the associated effective stored-energy function has been proposed by  \cite{APC16}. 
The required quasiconvexification of the principal solution---which is a formidable task, in general---may be accomplished in several special cases by showing that its rank-one convexification, which is analytically tractable, is polyconvex and, therefore, quasiconvex. 
This strategy, introduced by \cite{BJ87} in the context of phase transformation in shape-memory alloys, has been successfully employed by \cite{APC16} in computing the relaxation of the homogenization estimate by \cite{LPPC06} for the principal solution of a certain class of 2-D reinforced elastomers. 
In addition, making use of this strategy, \cite{FPC18} obtained the relaxation of the principal solution for the effective stored-energy function of two-phase Neo-Hookean laminates under plane-strain loading conditions.
An important conclusion of these works, which is particularly relevant to the present work, is that the relaxed or post-bifurcation states of these hyperelastic composites corresponds physically to the formation of lamellar domains.
In particular, the domains in these materials develop at a ``mesoscopic'' length scale, which is much smaller than the dimensions of the macroscopic specimen under consideration, but still much larger that the ``microscopic'' length scale of the heterogeneity.
It has also been found in these works that, in general, the principal solution loses (global) rank-one convexity before it loses strong ellipticity. 
However, for the special case when the applied loading is aligned with the principal axes of the microstructure in the pre-bifurcation state, it turns out that the material loses rank-one convexity and strong ellipticity simultaneously.
For this special, the solution obtained by \cite{FPC18} for the post-bifurcation behavior of Neo-Hookean laminates coincides, in effect, with the corresponding solution obtained earlier by \cite{SdATW16}. 
%

As already mentioned, in this work we investigate the effect of fiber plasticity on the macroscopic response and domain formation in soft biological composites.
A first effort to this end has already been made by \cite{FAPAB22}. However, these authors restricted their considerations in studying the critical conditions for the macroscopic LOE of a homogenization model for elastoplastic fiber-reinforced composites, derived on the basis of an approximation of the Voigt type. 
By contrast, in this work we make use of exact results for periodic elastoplastic laminates and compute the onset and evolution of domains in the post-bifurcation regime of these composites. 
For this purpose, we first discuss in section 2 the uniform-per-phase solution for the homogenized, incremental behavior of two-phase laminates with piecewise linear constituents at arbitrary finite strains.
In section 3, making use of this solution, we compute the principal and post-bifurcation behavior of laminates with elastoplastic constituents under plane-stain, axial loading along the direction of the layers. 
In this context, we derive a simple formula  for the macroscopic LOE of the composite, which implies that, just like their hyperelastic counterparts, elastoplastic laminates lose macroscopic ellipticity when their effective incremental strength in shear perpendicular to the layers vanishes for the first time. 
It is also shown that loss of macroscopic ellipticity is a necessary condition for a bifurcation of the long-wavelength type to occur, as well as that the post-bifurcation solution corresponds to the formation of lamellar domains at the mesoscopic level of these composites, just like those obtained by \cite{FPC18} for Neo-Hookean laminates. 
In section 4, in order to model and highlight the effect of fiber plasticity on the macroscopic response of soft biological composites, we specialize further the constitutive relations of section 3 to the case where only the elastically stiffer set of layers (fibers) admits plastic deformations, while the remaining layers (matrix) are taken to be purely hyperelastic. 
In this context, we study the macroscopic response and domain formation in these laminates under nonmonotonic loading conditions by means of specific numerical calculations. 
In the first part of section 4, we focus on the pre- and post-bifurcation response of these composites, while in the second part, we study in detail the influence of the material and loading parameters that are involved on the critical condition for macroscopic LOE. 
It is found that the plastic properties of the fibers have a significant effect on the onset of macroscopic LOE and a rather negligible effect on the post-bifurcation behavior of the composite. 
A detailed discussion concerning the incremental constitutive relations that are used to describe the local behavior in these composites at finite elastic and plastic strains is given in the Appendix. 
The bifurcation analysis in this article (Part I) is complemented by a corresponding imperfection analysis, presented in a forthcoming article (Part II), in the context of which the layers in the composite materials of interest are taken to have some small initial waviness.
%
%

\section{The homogenized behavior of laminates under general loading conditions}
\label{EB}

%
\begin{figure}
\centering
	\begin{tabular}{cc}
		\includegraphics[width=2.1in]{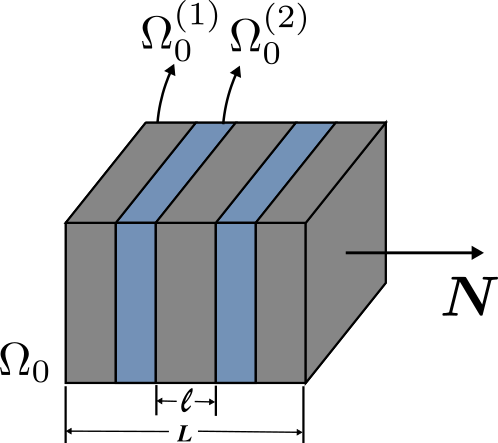}  \hspace{2 cm}
		& \includegraphics[width=2.1in]{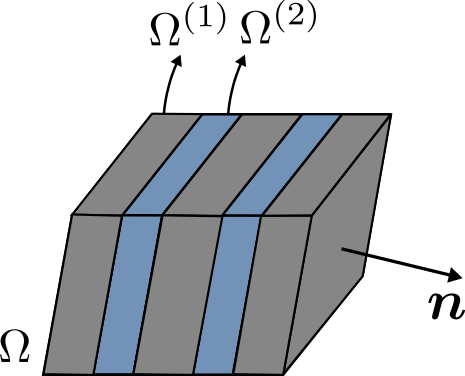}  \\
		(a) Undeformed state \hspace{2 cm} & (b) Deformed state \\
	\end{tabular}
	\caption{Schematic representation of a representative volume element $\Omega_0$ of a periodic, two-phase laminate, subjected to an affine deformation of the form (\ref{xhat}) on its boundary $\p \Omega_0$. (a) Undeformed configuration $\Omega_0$. (b) Deformed configuration $\Omega$.}
	\label{SchLam}
\end{figure}

Consider a ``representative volume element'' (RVE) $\Omega_0$ of a simple laminate, comprised of two different homogeneous phases that are layered periodically along a given direction $\bfN$, as shown in the schematic of Fig. \ref{SchLam}(a). 
The width of the unit cell and the overall dimensions of the RVE $\Omega_0$ are characterized by the ``microscopic'' and ``macroscopic'' length scales $\ell$ and $L$, respectively, which are assumed to be well-separated, i.e., $\ell << L$.
In addition, the constituent layers in the RVE are assumed to be perfectly bonded, so that the deformation and traction vector fields are continuous at the phase boundaries. The complementary parts of $\Omega_0$ occupied by the two types of layers are denoted by $\Omega^{(r)}_0$, with $r=1,2$. 
Letting $\bfX$ denote the position vector of the generic material point in $\Omega_0$ and $X = \bfX \cdot \bfN$,
the characteristic functions of the phases $\chi_0^{(r)}(X)$ are defined such that $\chi_0^{(r)}(X)=1$ if $\bfX$ is in $\Omega_0^{(r)}$ and $\chi_0^{(r)}(X)=0$, otherwise. 
These functions are required to satisfy the identity $\chi_0^{(1)}(X) + \chi_0^{(2)}(X) = 1$ at each point $\bfX$ in $\Omega_0$ and are, therefore, not independent. 
In what follows, we make use of the notation $\langle \cdot \rangle_0$ and $\langle \cdot \rangle_0^{(r)}$ for the volume averages of a scalar or tensor valued field over $\Omega_0$ and $\Omega_0^{(r)}$, respectively. 
Note that $\langle \chi_0^{(r)}(X) \rangle_0$ is equal to the volume fraction $c_0^{(r)} \equiv |\Omega_0^{(r)}| / |\Omega_0|$ of phase $r$ in $\Omega_0$.

The effective constitutive behavior of the composite material under consideration may be defined in terms of the average response of the RVE $\Omega_0$ under suitable loading paths that are compatible with macroscopically uniform fields (\cite{H63,H72}). Specifically, with no loss of generality, consider the current state $\Omega$ of the RVE at time $t$ obtained from its reference state $\Omega_0$ at time $t_0$ by an affine deformation of the form
\begin{equation}
	\bfx = \obfF \bfX \quad  
	\text{on} \quad  \p \Omega_0,
	\label{xhat}
\end{equation}
where $\p \Omega_0$ denotes the boundary of $\Omega_0$ and $\obfF = \obfF(t)$ is an arbitrary second-order tensor with positive determinant, which varies with $t$ but not with $\bfX$, serving to prescribe the loading path in strain space. It can be shown by using Gauss's theorem that 
\begin{equation}
	\langle \bfF \rangle_0 = \obfF,
	\label{Fave}
\end{equation}
where $\bfF = \bfF(\bfX, t)$ is the deformation gradient field produced by any continuous deformation field $\bfx = \bfx (\bfX, t)$ in $\Omega_0$ that satisfies the boundary condition (\ref{xhat}). Considering the rate form of (\ref{xhat}), i.e.,
\begin{equation}
	\dot{\bfx} = \dot{\obfF} \bfX \quad  
	\text{on} \quad  \p \Omega_0,
	\label{xdothat}
\end{equation}
it can be similarly shown that 
\begin{equation}
	\langle \dot{\bfF} \rangle_0 = \dot{\obfF},
	\label{Fdotave}
\end{equation}
and letting $\bfP = \bfP(\bfX, t)$ be any self-equilibrated first Piola-Kirchhoff stress field in $\Omega_0$, it can also be shown that the average of the local stress power $\mathcal{P} = \bfP \cdot \dot{\bfF}$ over the RVE $\Omega_0$ is given by
\begin{equation}
	\langle \bfP \cdot \dot{\bfF} \rangle_0 = \obfP \cdot \dot{\obfF}, \quad\quad \obfP \equiv \langle \bfP \rangle_0.
	\label{PBar}
\end{equation}
The result $(\ref{PBar})_1$, along with $(\ref{Fdotave})$ and $(\ref{PBar})_2$, shows that the first Piola-Kirchhoff stress $\bfP$ and the deformation gradient $\bfF$ constitute a pertinent conjugate pair of stress and strain measures for carrying out the transition from the microscopic level of the local material behavior in the layers to the macroscopic level of the homogenized constitutive behavior of the composite. Any other stress and strain measure may be defined at the macroscopic level in terms of $\obfP$ and $\obfF$, according to the standard theory of continuum mechanics.
For example, the macroscopic Cauchy stress tensor $\obfsig$ is defined by
\begin{equation}
	\obfsig = \oJ^{-1} \obfP \obfF^{T},
	\label{SigmaBar}
\end{equation}
where $\oJ = \det \obfF$. Using (\ref{xhat}), the volume average of the Cauchy stress field $\bfsig = \bfsig(\bfx, t)$ over $\Omega$ can be shown to coincide with its macroscopic counterpart, as defined by (\ref{SigmaBar}). It should be emphasized, however, that, in general, the pertinent volume average of a stress or strain measure is different than its macroscopic counterpart.

Next, we specialize our considerations by assuming that the local fields $\bfP = \bfP(\bfX, t)$ and $\bfF = \bfF(\bfX, t)$ in $\Omega_0$ are related via a piece-wise linear, incremental equation of the form
\begin{equation}
	\dot{\bfP} = \bmL (\bfX, t) \dot{\bfF}, \quad\quad 
	\bmL (\bfX, t) = \chi_0^{(1)}(X) \bmL^{(1)}(\bfX, t) + \chi_0^{(2)}(X) \bmL^{(2)}(\bfX, t),
	\label{LMBLam}
\end{equation}
where $\bmL^{(1)}(\bfX, t)$ and $\bmL^{(2)}(\bfX, t)$ are the associated incremental modulus tensors of the phases, which are assumed to depend on the deformation history at point $\bfX$ up to the current time $t$. The explicit dependence of these tensors on $\bfX$ has been introduced in order to emphasize the latter assumption and to account for the fact that, in general, the deformation history is different at different points, even within the same phase. Note that, this type of local material behavior includes as a special case the class of rate-independent, elastoplastic solids defined in the Appendix, for which the phase modulus tensors $\bmL^{(r)}$, with $r=1,2$, are given by expressions of the form $(\ref{Sdot})_2$.

Thus, assuming that the fields $\bmL (\bfX, t)$, $\bfP = \bfP(\bfX, t)$, and $\bfF = \bfF(\bfX, t)$ are known at the current time $t$, the rate fields $\dot{\bfP} = \dot{\bfP}(\bfX, t)$ and $\dot{\bfF} = \dot{\bfF}(\bfX, t)$ at time $t$ may be determined, in principle, by solving the incremental problem defined by the boundary condition $(\ref{xdothat})$, the local constitutive relation $(\ref{LMBLam})$, the rate-form of the equilibrium equations within the phases, neglecting body forces, and the continuity conditions for the deformation- and traction-rate fields on the phase boundaries. The solution of this problem allows in turn the determination of the effective, incremental constitutive relation for the laminate, i.e., the relation between $\dot{\obfP} = \langle \dot{\bfP} \rangle_0$ and $\dot{\obfF} = \langle \dot{\bfF} \rangle_0$, the integration of which over a given loading path yields the corresponding relation between the macroscopic stress $\obfP$ and deformation gradient $\obfF$. In general, the incremental problem for the fields $\dot{\bfP} = \dot{\bfP}(\bfX, t)$ and $\dot{\bfF} = \dot{\bfF}(\bfX, t)$ is difficult to solve analytically. However, for the special case when the these fields are uniform (but different) in each phase, this problem reduces to a set of algebraic equations, which can be solved explicitly, as detailed in the following subsection.

\subsection{Uniform-per-phase fields} 
\label{PUF}

We consider next the case that, at any given time $t$ during a loading process, the tensor-valued fields $\bmL = \bmL(\bfX, t)$, $\bfF = \bfF(\bfX,t)$, and $\bfP = \bfP(\bfX,t)$ are uniform in each phase domain $\Omega_0^{(r)}$, with $r=1,2$, and let $\bmL^{(r)}$, $\obfF^{(r)} \equiv \langle \bfF \rangle_0^{(r)}$, and $\obfP^{(r)} \equiv \langle \bfP \rangle_0^{(r)}$ be the (constant) values of these fields in phase $r$. Note that, the tensors $\bmL^{(r)} = \bmL^{(r)}(t)$, $\obfF^{(r)} = \obfF^{(r)}(t)$, and $\obfP^{(r)} = \obfP^{(r)}(t)$ may vary with time and depend, in general, on the applied loading path up to the current time $t$. At any given instant $t$ during the loading, the tensors $\obfF^{(r)}$ and $\obfP^{(r)}$ are required to satisfy the volume average identities
\begin{equation}
	c^{(1)}_0 \obfF^{(1)} + c^{(2)}_0 \obfF^{(2)} = \obfF, \quad\quad 
	c^{(1)}_0 \obfP^{(1)} + c^{(2)}_0 \obfP^{(2)} = \obfP,
	\label{AVEIDS}
\end{equation}
as well as the equations
\begin{equation}
	\obfF^{(2)} - \obfF^{(1)} = \bfA \otimes \bfN, \quad\quad
	\obfP^{(2)} \bfN = \obfP^{(1)} \bfN = \obfP \bfN,
	\label{PBConds}
\end{equation}
where $\bfA = \bfA(t)$ in $(\ref{PBConds})_1$ is an unknown vector, which depends on $t$, but not on $\bfX$. 
Equations $(\ref{PBConds})_1$ and $(\ref{PBConds})_2$ follow directly from the deformation and traction continuity conditions at the boundaries of the layers, while equation $(\ref{PBConds})_3$ is obtained by taking the product of $(\ref{AVEIDS})_2$ with $\bfN$ and making use of $(\ref{PBConds})_2$.

Under these considerations, the phase volume fractions $c^{(r)}$ and the lamination-orientation vector $\bfn$ in the current configuration  $\Omega$ of the composite can be shown (\cite{LPPC2009}) to be given by
\begin{equation}
	c^{(r)}  = \frac{ \det \obfF^{(r)} }{ \det \obfF } c_0^{(r)}, \quad\quad
	 \bfn = \frac{ \obfF^{-T} \bfN }{ |\obfF^{-T} \bfN| },
	\label{MSevol}
\end{equation}
respectively. In addition, conditions $(\ref{AVEIDS})_2$, $(\ref{PBConds})_2$, and $(\ref{PBConds})_3$ may be expressed in the current configuration as 
\begin{equation}
	c^{(1)} \obfsig^{(1)} + c^{(2)} \obfsig^{(2)} = \obfsig, \quad\quad
	\obfsig^{(2)} \bfn = \obfsig^{(1)} \bfn = \obfsig \bfn,
	\label{AVEID_TCC_C}
\end{equation}
where $\obfsig$ is the macroscopic Cauchy stress tensor, given by (\ref{SigmaBar}), and 
\begin{equation}
	\obfsig^{(r)} = \oJ^{(r)-1} \obfP^{(r)} \obfF^{(r)T},
	\label{SigmaBarr}
\end{equation}
with $\oJ^{(r)} = \det \obfF^{(r)}$, are the (constant) Cauchy stress tensors in the constituent phases of the laminate.

Thus, the incremental problem for the homogenized behavior of the laminate is defined by the rate form of the equations (\ref{AVEIDS}) and (\ref{PBConds}), i.e.,
\begin{equation}
	c^{(1)}_0 \dot{\obfF}^{(1)} + c^{(2)}_0 \dot{\obfF}^{(2)} = \dot{\obfF}, \quad\quad 
	c^{(1)}_0 \dot{\obfP}^{(1)} + c^{(2)}_0 \dot{\obfP}^{(2)} = \dot{\obfP}
	\label{AVEIDSRF}
\end{equation}
and 
\begin{equation}
	\dot{\obfF}^{(2)} - \dot{\obfF}^{(1)} = \dot{\bfA} \otimes \bfN, \quad\quad
	\dot{\obfP}^{(2)} \bfN = \dot{\obfP}^{(1)} \bfN = \dot{\obfP} \bfN,
	\label{PBCondsRF}
\end{equation}
respectively, along with the local constitutive relation (\ref{LMBLam}), which in this case takes the form 
\begin{equation}
	\dot{\obfP}^{(1)} = \bmL^{(1)} \dot{\obfF}^{(1)}, \quad\quad 
	\dot{\obfP}^{(2)} = \bmL^{(2)} \dot{\obfF}^{(2)},
	\label{CEqnPrRF}
\end{equation}
where we emphasize once again that the phase modulus tensors $\bmL^{(1)}$ and $\bmL^{(2)}$ depend on the loading path up to the current time, but are independent of $\bfX$. 
We also recall that, for the class of elastoplastic laminates of interest in this work, these tensors are defined by expressions of the form $(\ref{Sdot})_2$, given in the Appendix.

The local rate field $\dot{\bfA}$, $\dot{\obfF}^{(r)}$, and $\dot{\obfP}^{(r)}$ may be eliminated from the above equations $(\ref{AVEIDSRF})$, $(\ref{PBCondsRF})$, and $(\ref{CEqnPrRF})$, as follows.
First, making use of $(\ref{AVEIDSRF})_1$ and $(\ref{PBCondsRF})_1$, we find that 
\begin{equation}
	\dot{\obfF}^{(1)} = \dot{\obfF} - c^{(2)}_0 \dot{\bfA} \otimes \bfN, \quad\quad 
	\dot{\obfF}^{(2)} = \dot{\obfF} + c^{(1)}_0 \dot{\bfA} \otimes \bfN.
	\label{FBARrRF}
\end{equation}
Then, combining expressions $(\ref{FBARrRF})_1$, $(\ref{CEqnPrRF})_1$, and $(\ref{PBCondsRF})_3$, we obtain a linear equation for $\dot{\bfA}$, the solution of which is
\begin{equation}
	\dot{\bfA} = \frac{1}{c^{(2)}_0} \left(\bfK^{(1)}\right)^{-1} \left( \bmL^{(1)}  \dot{\obfF} - \dot{\obfP} \right) \bfN,
	\label{Adot}
\end{equation}
where $\bfK^{(1)}$ is the acoustic tensor associated with the modulus $\bmL^{(1)}$ and it is defined by
\begin{equation}
	K^{(1)}_{ik} = \mathcal{L}^{(1)}_{ijkl} N_j N_l.
	\label{KTens}
\end{equation}
It follows from $(\ref{Adot})$ that
\begin{equation}
	\dot{\bfA} \otimes \bfN = \frac{1}{c^{(2)}_0} \bmH^{(1)} \left( \bmL^{(1)}  \dot{\obfF} - \dot{\obfP} \right),
	\label{ANDyadic}
\end{equation}
where, for convenience, we have introduced the fourth-order tensor $\bmH^{(1)}$ with components 
\begin{equation}
	\mathcal{H}^{(1)}_{ijkl} = \left( \bfK^{(1)}\right)^{-1} _{ik} N_j N_l.
	\label{HTens}
\end{equation}
Finally, making use of expressions $(\ref{FBARrRF})$, along with the result $(\ref{ANDyadic})$, in the constitutive equations $(\ref{CEqnPrRF})$, and substituting the resulting expressions in the volume average identity $(\ref{AVEIDSRF})_2$, it can be shown that 
\begin{equation}
	\dot{\obfP} = \widetilde{\bmL} \, \dot{\obfF}, \quad\quad 
	\widetilde{\bmL} = \bmL^{(1)} + c^{(2)}_0 \left[ \left( \bmL^{(2)} - \bmL^{(1)} \right)^{-1} + c^{(1)}_0 \bmH^{(1)} \right]^{-1},
	\label{SFRF}
\end{equation}
which constitutes the desired incremental constitutive relation for the homogenized behavior of the laminate. 
The deformation-gradient rates in the constituent phases of the composite are then given by
\begin{equation}
	\dot{\obfF}^{(1)} = \bmA^{(1)} \dot{\obfF}, \quad\quad
	\dot{\obfF}^{(2)} = \bmA^{(2)} \dot{\obfF}
	\label{FBARrRF2}
\end{equation}
where 
\begin{equation}
	\bmA^{(1)} = \bmI + \bmH^{(1)} \left( \widetilde{\bmL} - \bmL^{(1)} \right), \quad\quad 
	\bmA^{(2)} = \bmI - \frac{c^{(1)}_0}{c^{(2)}_0} \bmH^{(1)} \left( \widetilde{\bmL} - \bmL^{(1)} \right),
	\label{ArTens}
\end{equation}
with $\bmI$ being the fourth-order identity tensor with components $\mathcal{I}_{ijkl} = \delta_{ik} \delta_{jl}$. The results $(\ref{FBARrRF2})$, along with $(\ref{ArTens})$, follow by substituting $(\ref{SFRF})_1$ in $(\ref{ANDyadic})$ and using the resulting expression for $\dot{\bfA} \otimes \bfN$ in $(\ref{FBARrRF})$.

In summary, the response of the laminate along any prescribed loading path may be determined by integrating the system of incremental equations $(\ref{SFRF}) - (\ref{ArTens})$ along the given path. 
Note, however, that the phase modulus tensors $\bmL^{(r)}$ in the context of these equations are given as functions of the local deformation gradients $\obfF^{(r)}$ and, therefore, equations $(\ref{SFRF}) - (\ref{ArTens})$ are nonlinear and their integration should be pursued numerically, in general.

We conclude this section with a few remarks concerning the application of the piece-wise uniform solution $(\ref{SFRF}) - (\ref{ArTens})$ to the class of rate-independent, elastoplastic laminates of interest in this work. 
As already mentioned, the phase modulus tensors $\bmL^{(r)}$ for this class of materials are given by expressions of the form $(\ref{Sdot})_2$ (Appendix).
In the linear elastic regime of these elastoplastic composites, the variables $\bmL^{(r)}$ reduce to the corresponding ground-state stiffness tensors, which are known to be constant and positive definite. Hence, in this regime, the piece-wise uniform solution $(\ref{SFRF}) - (\ref{ArTens})$ is explicit and can be shown to be unique and stable.
The response of the material along any prescribed loading path which originates in the undeformed state and gets into the finite-strain regime may be determined by integrating equations $(\ref{SFRF}) - (\ref{ArTens})$, starting from the unique solution in the linear elastic regime. 
Note that, these equations are coupled with the constitutive relations given in the Appendix for the elastoplastic constituents and, therefore, the local elastic and plastic strains as well as the stresses in the constituent phases of the laminate are also determined as a byproduct of the integration procedure (see subsection \ref{NI}). 
The solution for the local and macroscopic behavior of the laminate obtained in this way is referred to as the ``principal solution'' (PS).
In this connection, it is important to emphasize that the principal solution is not necessarily unique for arbitrary loading paths. 
In fact, as shown by \cite{TM85} (see also \cite{GMT93} and \cite{NT04}), there exist loading paths along which the layered composites of interest reach a critical state at which the principal solution may bifurcate to a different mode, characterized by either a short or a long (compared with the length scale $\ell$ of the heterogeneity) wavelength. 
It has also been shown by these authors that the onset of bifurcations of the long-wavelength type coincides with the loss of strong ellipticity of the incremental equations of equilibrium for the homogenized behavior of the laminate, which takes place when the condition 
\begin{equation}
	\det \widetilde{\bfK} = 0, \quad\quad 
	\widetilde{K}_{ik} = \widetilde{\mathcal{L}}_{ijkl} V_j V_l,
	\label{LOECond}
\end{equation}
is first met for some unit vector $\bfV$, where we recall that $\widetilde{\bmL}$ is the incremental effective modulus tensor of the laminate, given by $(\ref{SFRF})_2$. 
In this regard, it is also important to recall from the above-referenced work that under plane-strain loading conditions such that the principal axes of the loading are aligned (or nearly aligned) with those of the microstructure, which are particularly relevant to soft-biological composites, like tendons, the critical mode---i.e., the first bifurcation that occurs as the material is monotonically loaded from the undeformed state---is always of the long-wavelength type, provided that the concentration of the stiffer phase is not too small (typically, when it is greater than $10\%$). Based on these findings, in this work we focus our consideration on bifurcations of the long-wavelength type, as determined by the macroscopic loss of ellipticity (LOE) condition $(\ref{LOECond})$. 

\section{The plane-strain response of laminates under aligned loading conditions}
\label{EPLamsAPSL}

\begin{figure}
	\begin{tabular}{ccc}
		\centering
		\includegraphics[width=2.in]{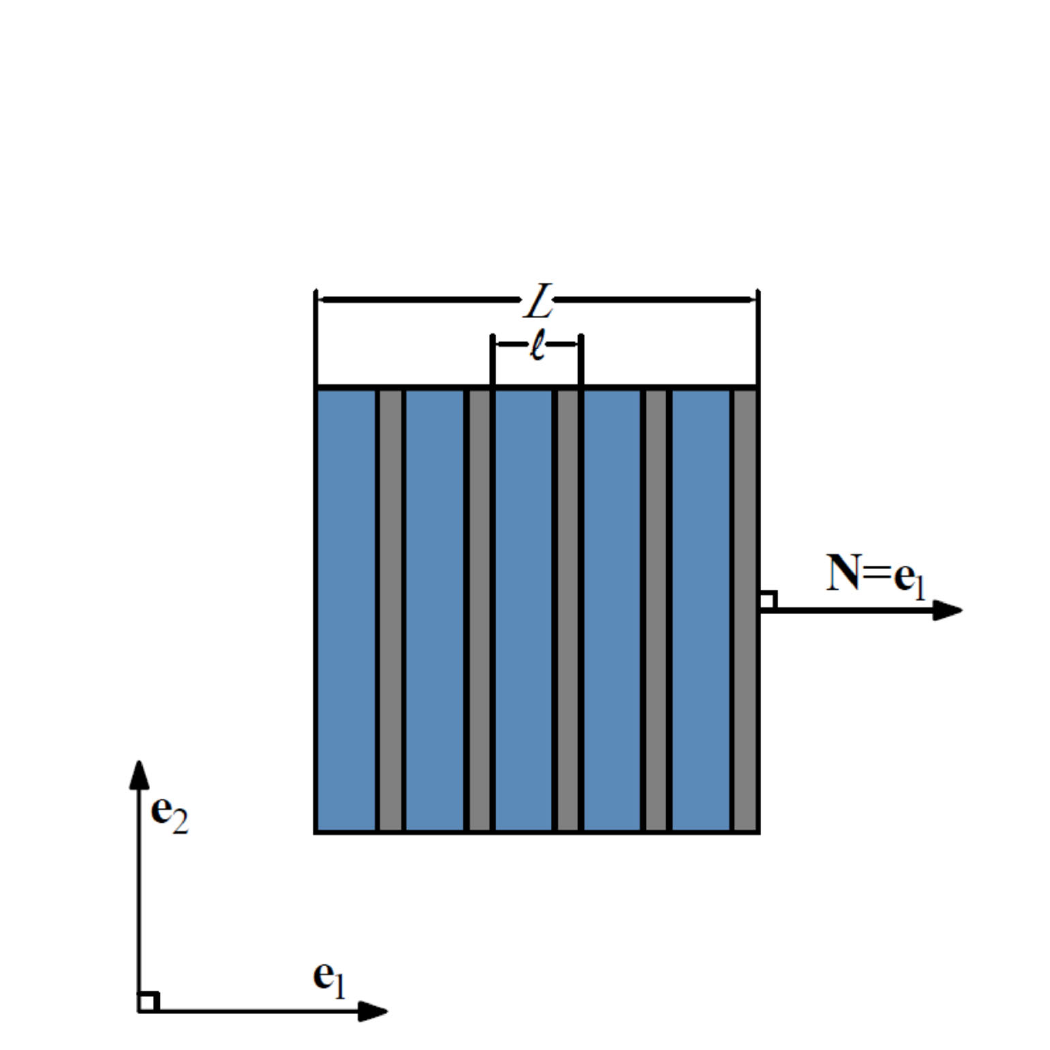} 
		& \includegraphics[width=2.in]{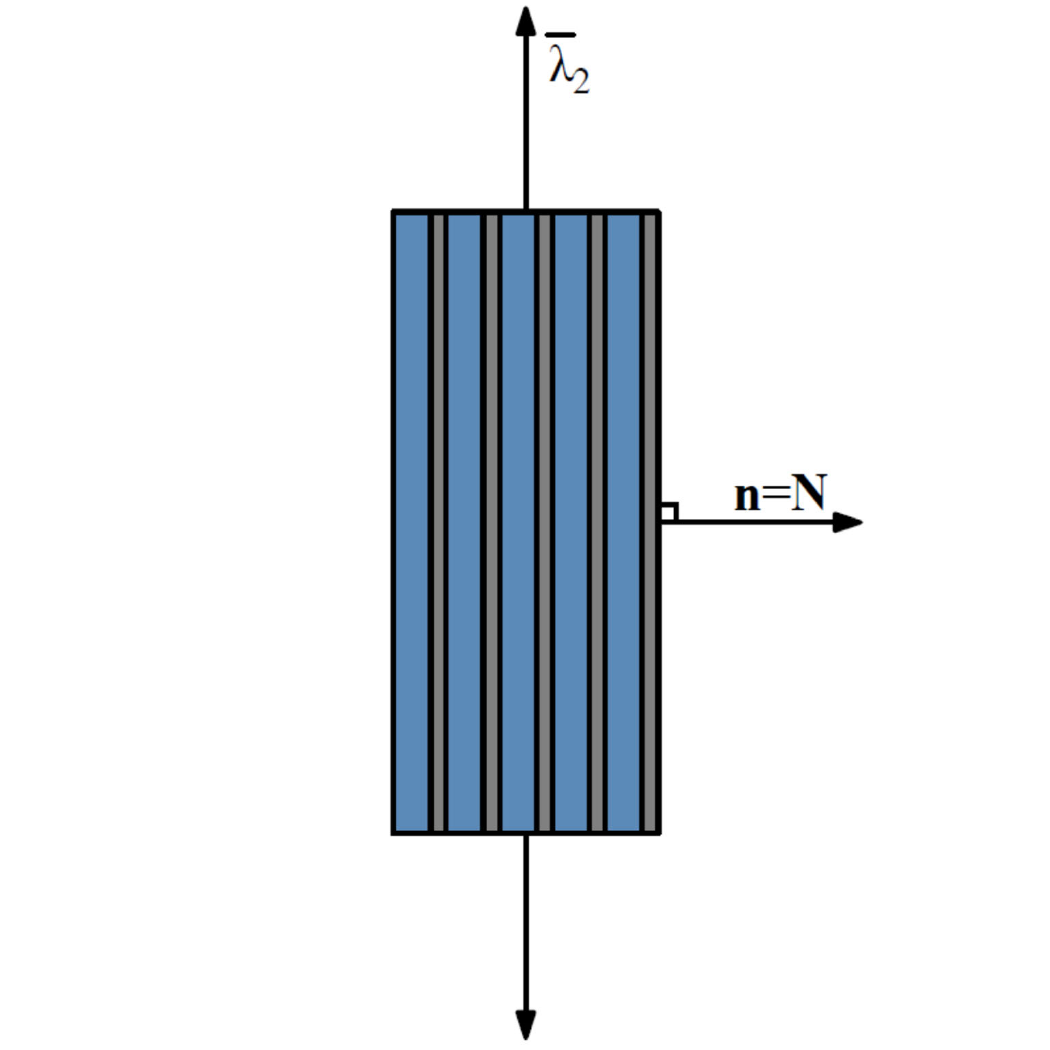} & \includegraphics[width=2.in]{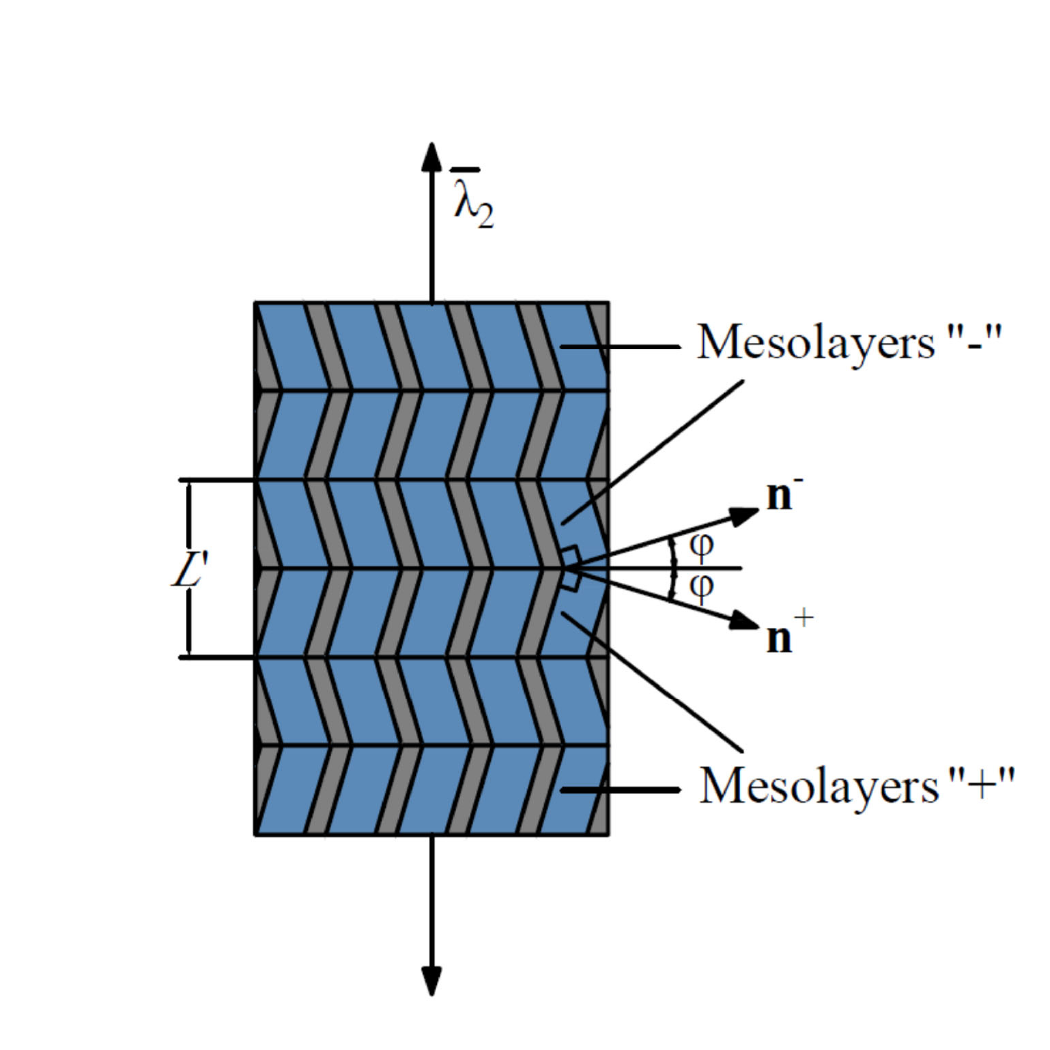} \\
		(a) Undeformed state & (b) Deformed principal state & (c) Deformed post-bifurcation state \\
	\end{tabular}
	\caption{Schematic representation of a representative volume element (RVE) of a two-phase laminate subjected to a unidirectional loading along $\bfe_2$, with the deformation confined on the plane $\bfe_1-\bfe_2$. (a) The undeformed state ($\ol_2 = 1$). (b) A deformed state ($\ol_2 > 1$) along the principal path. (c) A deformed state ($\ol_2 > 1$) along the post-bifurcation path.}
	\label{Sch_AL}
\end{figure}
%

%
In this section we specialize the general results of the previous section to laminates comprised of elastoplastic
constituents and subjected to plane-strain loading conditions.
In addition, motivated by applications in soft biological composites, we restrict our considerations to loadings that are aligned with the direction of the layers on the deformation plane.

%
Let an orthonormal set of vectors $\bfe_i$ ($i=1,2,3$) define a fixed coordinate frame, with $\bfe_1 = \bfN$ and $\bfe_2$ lying on the deformation plane,
as shown in Fig. \ref{Sch_AL}(a). 
In what follows, we refer the components of all vectors and tensors to the fixed Cartesian system with basis vectors $\bfe_i$. We restrict our considerations to the special case that the material is subjected macroscopically to a unidirectional loading along $\bfe_2$, so that the macroscopic deformation gradient, first Piola-Kirchhoff, and Cauchy stress tensors have the following matrix forms
\begin{equation}
	\left[ \oF \right] = \text{diag}\{\ol_{1}, \ol_{2}, 1\}, \quad\quad 
	\left[ \oP \right] = \text{diag}\{0, \oP_{2}, \oP_{3}\}, \quad\quad 
	\left[ \osig \right] = \text{diag}\{0, \osig_{2}, \osig_{3}\}
	\label{AL_MacroFields}
\end{equation}
respectively, where the notation ``$\text{diag}$'' indicates the diagonal components of a matrix in the orthonormal basis $\bfe_i$.
It follows from the more general relation (\ref{SigmaBar}) that the unknown components in (\ref{AL_MacroFields}) are related by
\begin{equation}
	\oP_2 = \ol_1 \osig_2 \quad\quad 
	\oP_3 = \ol_1 \ol_2 \osig_3.
	\label{P2P3}
\end{equation}
Next, letting $\ol_{2}$ be the (prescribed) loading parameter, we make use of the constitutive relations for the local and macroscopic behavior of the laminate in order to determine the variables $\ol_{1}$, $\osig_{2}$, and $\osig_{3}$ in terms of $\ol_{2}$. The details of this calculation in the pre- and post-bifurcation regime are given in the following subsections \ref{PSol} and \ref{PBSol}, respectively.

\subsection{Principal solution}
\label{PSol}

To simplify the analysis, we first anticipate that the principal solution for the deformation gradient and Cauchy stress tensors in the constituent phases of the laminate have the following matrix forms 
\begin{equation}
	\left[ \oF^{(r)} \right] = \text{diag}\{\ol^{(r)}_1, \ol^{(r)}_2, \ol^{(r)}_3\}, \quad\quad
	\left[ \osig^{(r)} \right] = \text{diag}\{\osig^{(r)}_1, \osig^{(r)}_2, \osig^{(r)}_3\},
	\label{AL_MicroFields}
\end{equation}
respectively, with $r=1,2$. Thus, the volume average identities $(\ref{AVEIDS})_1$ and $(\ref{AVEID_TCC_C})_1$ for these tensors reduce to
\begin{equation}
	c_0^{(1)} \ol^{(1)}_i +c_0^{(2)} \ol^{(2)}_i = \ol_i, \quad\quad
	c^{(1)} \osig^{(1)}_i +c^{(2)} \osig^{(2)}_i = \osig_i, \quad\quad i=1,2,3,
	\label{AL_VolAveIDs}
\end{equation}
where $\ol_3 = 1$ and $\osig_1 = 0$, and where it is recalled that $c_0^{(r)}$ and 
$c^{(r)}$
are the volume fractions of the phases in the reference and current configuration, respectively. In addition, the continuity conditions $(\ref{PBConds})_1$ and $(\ref{AVEID_TCC_C})_2$ for the deformation and traction vectors on the phase boundaries of the composite reduce to
\begin{equation}
	\ol^{(r)}_2 = \ol_2, \quad\quad
	\ol^{(r)}_3 = \ol_3 = 1,
	\label{AL_DCC}
\end{equation}
and
\begin{equation}
	\osig^{(r)}_{1} = 0, 
	\label{AL_TCC}
\end{equation}
respectively. Moreover, the multiplicative decomposition $\obfF^{(r)} = \obfF^{(r,e)} \obfF^{(r,p)}$ for each phase $r$ may be expresses as 
\begin{equation}
	\ol^{(r)}_i =\ol^{(r,e)}_i \ol^{(r,p)}_i, \quad\quad i=1,2,3,
	\label{AL_Lr}
\end{equation}
where $\ol^{(r,e)}_i$ and $\ol^{(r,p)}_i$ are respectively the elastic and plastic parts of the total stretch $\ol^{(r)}_i$. Making use of the compatibility conditions (\ref{AL_DCC}), it is convenient to rewrite the decomposition $(\ref{AL_Lr})$ for $i=2$ and $i=3$, along with the plastic incompressibility constraint $\ol^{(r,p)}_1 \ol^{(r,p)}_2 \ol^{(r,p)}_3 = 1$, in the following form
\begin{equation}
	\ol^{(r,e)}_2 = \frac{\ol_2}{\ol^{(r,p)}_2}, \quad\quad
	\ol^{(r,e)}_3 = \ol^{(r,p)}_1 \ol^{(r,p)}_2, \quad\quad
	\ol^{(r,p)}_3 = \frac{1}{\ol^{(r,p)}_1 \ol^{(r,p)}_2},
	\label{AL_LEr2LEr3LPr3}
\end{equation}
thus, reducing the unknown stretches in the phases to $\ol^{(r,e)}_1$, $\ol^{(r,p)}_1$, and $\ol^{(r,p)}_2$. These unknown, together with the accumulated plastic strain $\oeps^{(r,p)}$ in each phase, are determined in terms of the loading parameter $\ol_{2}$ by making use of the traction continuity condition  $(\ref{AL_TCC})$, along with the constitutive equations of the phases, as detailed in the sequel.

For simplicity, we restrict further our considerations to the special case that the elastic part of the local material behavior is described by the Neo-Hookean stored-energy function
\begin{equation}
	W^{(r)} \left( \obfF^{(r,e)}  \right)= \frac{\mu^{(r)}}{2} \left(\oI^{(r,e)} -3 -2\ln \oJ^{(r,e)} \right) + \frac{\beta^{(r)}}{2} \left( \oJ^{(r,e)}-1 \right)^2, \quad\quad
	\beta^{(r)} = \kappa^{(r)} - \frac{2 \mu^{(r)}}{3},
	\label{WrNH}
\end{equation}
where $\oI^{(r,e)} = \obfF^{(r,e)} \cdot \obfF^{(r,e)}$, $\oJ^{(r,e)} = \det \obfF^{(r,e)}$, and $\beta^{(r)}$, $\mu^{(r)}$, and $\kappa^{(r)}$ are respectively the Lame, shear, and bulk modulus of phase $r$ at zero strain. For this special case, it can be easily shown that 
\begin{equation}
	\osig^{(r)}_{i} = \mu^{(r)} \frac{ \left(\ol^{(r,e)}_{i} \right)^2 - 1 }{  \ol^{(r,e)}_1 \ol_2 \ol^{(r,p)}_1  } + 
	\beta^{(r)} \left(  \ol^{(r,e)}_1 \ol_2 \ol^{(r,p)}_1  - 1 \right), \quad\quad i=1,2,3.
	\label{AL_Sr}
\end{equation}
It can also be shown that the Mandel stress tensor $\obfSig^{(r)}$ in each phase $r$ is equal to the corresponding Kirchhoff stress  $\obftau^{(r)} = \oJ^{(r,e)} \obfsig^{(r)}$. Thus, the associated equivalent stress measure $\oSig^{(r)}_{eq}$ that enters the definition of the von Mises yield criterion for phase $r$ takes the form
\begin{equation}
	\oSig^{(r)}_{eq} = \Bigg{\{} \frac{3}{2} \left[ \left(\oSig^{(r,d)}_1 \right)^2 + \left(\oSig^{(r,d)}_2 \right)^2 + \left(\oSig^{(r,d)}_3 \right)^2 \right] \Bigg{\}}^{1/2},
	\label{AL_YFcn2}
\end{equation}
where $\oSig^{(r,d)}_i$ are the principal components of the deviatoric part of $\obfSig^{(r)}$, given by
\begin{equation}
	\oSig^{(r,d)}_1 = \frac{ \mu^{(2)} } {3} \left[2 \left( \ol^{(r,e)}_{1} \right)^2 - \left( \ol_{2} / \ol^{(r,p)}_{2} \right)^2 - \left( \ol^{(r,p)}_1 \ol^{(r,p)}_2 \right)^2 \right], \quad
	\oSig^{(r,d)}_2 = \frac{ \mu^{(2)} } {3} \left[2 \left( \ol_{2} / \ol^{(r,p)}_{2} \right)^2 - \left( \ol^{(r,e)}_1 \right)^2 - \left( \ol^{(r,p)}_1 \ol^{(r,p)}_2 \right)^2 \right], 
	\label{AL_MandelSigmar}
\end{equation}
and $\oSig^{(r,d)}_3 = - \oSig^{(r,d)}_1 - \oSig^{(r,d)}_2$.

Given expression $(\ref{AL_Sr})$ for the stress component $\osig^{(r)}_{1}$, the traction continuity condition $(\ref{AL_TCC})$ defines a quadratic equation for $\ol^{(r,e)}_1$, the positive root of which is given by
\begin{equation}
	\ol^{(r,e)}_1 = \frac{ \beta^{(r)} \ol_2 \ol^{(r,p)}_1 + \sqrt{ \left(\beta^{(r)} \ol_2 \ol^{(r,p)}_1 \right)^2 + 4 \mu^{(r)} \left[ \mu^{(r)} + \beta^{(r)} \left( \ol_2 \ol^{(r,p)}_1 \right)^2 \right]} }{2  \left[ \mu^{(r)} + \beta^{(r)} \left( \ol_2 \ol^{(r,p)}_1 \right)^2 \right]}.
	\label{AL_LE1r}
\end{equation}
Thus, the unknown variables reduce to $\ol^{(r,p)}_1$, $\ol^{(r,p)}_2$, and $\oeps^{(r,p)}_2$. These variables are treated as the principal unknowns and are obtained from the consistency condition and the components of the normality rule in directions 1 and 2, which read respectively as
\begin{equation}
	\dot{\Sigma}^{(r)}_{eq} = \frac{d \Sigma^{(r)}_y}{d \oeps^{(r,p)}} \dot{\oeps}^{(r,p)}, \quad\quad
	\frac{\dot{\l}^{(r,p)}_1}{\l^{(r,p)}_1} = \frac{3 \dot{\oeps}^{(r,p)}}{2 \Sigma^{(r)}_y} \Sigma^{(r,d)}_1, \quad\quad 
	\frac{\dot{\l}^{(r,p)}_2}{\l^{(r,p)}_2} = \frac{3 \dot{\oeps}^{(r,p)}}{2 \Sigma^{(r)}_y} \Sigma^{(r,d)}_2,  
	\label{AL_ODEs}
\end{equation}
where $\Sigma^{(r)}_y$ is the current yield stress of phase $r$ and the superimposed dot on a variable indicates the derivative of that variable with respect to $\ol_2$.
The above expressions $(\ref{AL_ODEs})$ constitute a nonlinear system of ODEs for the variables $\ol^{(r,p)}_1$, $\ol^{(r,p)}_2$, and $\oeps^{(r,p)}_2$, which must be integrated numerically along any portion of the loading path involving plastic deformations. For any part of the loading path involving only elastic deformations, the variables $\ol^{(r,p)}_{1}$, $\ol^{(r,p)}_{2}$, and $\oeps^{(r,p)}$ are fixed and equations $(\ref{AL_LE1r})$ determine the elastic stretches $\ol^{(r,e)}_1$ explicitly in terms of $\ol_2$. In any case, given the stretches $\ol^{(r,p)}_1$, $\ol^{(r,p)}_2$, and $\ol^{(r,e)}_1$, the local and macroscopic response of the laminate may be readily obtained from $(\ref{AL_Sr})$ and $(\ref{AL_VolAveIDs})$. 
As can be seen from these results, due to special type of loading under consideration here, the stress-stretch response of each constituent phase of the laminate is independent from that of the other phase. 

Finally, under the considerations of this subsection, it turns out that the condition $(\ref{LOECond})$ for the loss of macroscopic ellipticity of the elastoplastic laminate takes the simple form
\begin{equation}
	\tilde{\cal{L}}_{1212} =0, \quad\quad
	\tilde{\cal{L}}_{1212} = c_0^{(1)} \frac{ \mu^{(1)} }{ (\ol^{(1,p)}_2)^2 } + c_0^{(2)} \frac{ \mu^{(2)} }{ (\ol^{(2,p)}_2)^2 } - 
	\frac{1}{\ol_2^2} \frac{ c_0^{(1)} c_0^{(2)} \left( \mu^{(2)} \ol_1^{(2,e)} / \ol_1^{(2,p)} - \mu^{(1)} \ol_1^{(1,e)} / \ol_1^{(1,p)} \right)^2 }{c_0^{(1)} \mu^{(2)} / (\ol_1^{(2,p)})^2 + c_0^{(2)} \mu^{(1)} / (\ol_1^{(1,p)})^2 }.
	\label{AL_LOEcond}
\end{equation}
The above result $(\ref{AL_LOEcond})_1$ implies that the laminate loses macroscopic ellipticity when its incremental strength in shear perpendicular to the layers, as measured by the effective modulus $\tilde{\cal{L}}_{1212}$, vanishes for the first time. 
It should be emphasized that, despite its appearance, the LOE condition $(\ref{AL_LOEcond})$ depends on both the loading path and the plastic properties of the constituent phases; this dependence enters through the dependence of the plastic stretches $\ol_1^{(r,p)}$ and $\ol_2^{(r,p)}$ on the latter factors and it is, therefore, implicit.
For the special case of a laminate with purely elastic phases, i.e., when $\ol_1^{(r,p)}=\ol_2^{(r,p)}=1$ for all values of $\ol_2$, expression $(\ref{AL_LOEcond})_2$ reduces to
\begin{equation}
	\tilde{\cal{L}}_{1212} = c_0^{(1)} \mu^{(1)} + c_0^{(2)} \mu^{(2)} - 
	\frac{1}{\ol_2^2} \frac{ c_0^{(1)} c_0^{(2)} \left( \mu^{(2)} \ol_1^{(2)} - \mu^{(1)} \ol_1^{(1)} \right)^2 }{c_0^{(1)} \mu^{(2)} + c_0^{(2)} \mu^{(1)}},
	\label{AL_LOEcond_E}
\end{equation}
where the purely elastic stretches $\ol_1^{(r)} = \ol_1^{(r,e)}$ are given explicitly in terms of the applied stretch $\ol_2$ by expression $(\ref{AL_LE1r})$ for $\ol_1^{(r,p)}=1$.
It is recalled that, when the macroscopic LOE condition $(\ref{AL_LOEcond})$ (or $(\ref{AL_LOEcond_E})$, if pertinent) is first met along the principal equilibrium path, a bifurcation to a long-wavelength mode becomes possible for the first time.

\subsection{Post-bifurcation solution}
\label{PBSol}

Next, we discuss the long-wavelength post-bifurcation solution for the class of elastoplastic laminates under the aligned loading conditions of interest. 
In particular, we search for a (non isochoric) solution that has (otherwise) the same form as the corresponding solution obtained by \cite{FPC18} for laminates with incompressible Neo-Hookean constituents.
The main features of this solution are illustrated schematically in Fig. \ref{Sch_AL}(c), which shows the corresponding deformed, post-bifurcation configuration of the material.
The state shown in Fig. \ref{Sch_AL}(c) is characterized by periodic lamellar domains that are formed at a ``mesoscopic'' length-scale $L'$ which is much larger than the ``microscopic'' length-scale $\ell$ of the original unit cell, but still much smaller than the ``macroscopic'' length-scale $L$ of the RVE $\Omega_0$, i.e., $L >> L' >> \ell$.
Thus, at its generic post-bifurcation state, the composite exhibits fine lamellar microstructures at two well-seperated length-scales and it is, therefore, a rank-2 laminate.
In what follows, the lamellar domains at the length-scale $L'$ are referred to as ``mesolayers,'' while their counterparts at the length-scale $\ell$ are termed ``microlayers.''
Furthermore, we make use of the labels ``+'' and ``-'' in order to distinguish between the two different types of mesolayers, while still using the labels ``1'' and ``2'' for the two different types of microlayers. 
The deformation gradient and first Piola-Kirchhoff stress fields are taken to be piece-wise uniform both at the level of microlayers and at the level of mesolayers, whereas the homogenized behavior of the mesolayers and the macroscopic, post-bifurcation behavior of the
rank-2 laminate are determined by making use of the piece-wise uniform solution of subsection \ref{PUF}, as detailed in the sequel.

%
%

Based on the aforementioned solution of \cite{FPC18}, we first anticipate that the volume fraction of each type of mesolayer  is equal to $1/2$, while the corresponding lamination direction (i.e., the direction normal to the mesolayers) is 
parallel to the macroscopic loading direction $\bfe_2$, as shown in Fig. \ref{Sch_AL}(c). In addition, the deformation gradient and first Piola-Kirchhoff stress tensors in the mesolayers are taken to be of the forms
\begin{equation}
	\left[ \oF^{\pm} \right] =
	\begin{bmatrix}
		\ol_1 & \pm \ogamma^* & 0 \\
		0 & \ol_2 & 0 \\
		0 & 0 & 1
	\end{bmatrix}, \quad\quad
	\left[ \oP^{\pm} \right] =
	\begin{bmatrix}
		0 & 0 & 0 \\
		\pm \oG^*  & \oP_{2} & 0 \\
		0 & 0 & \oP_{3}
	\end{bmatrix},
	\label{PBS_MesoFields}
\end{equation}
where $\ogamma^*$ and $\oG^*$ are the corresponding amounts of shear deformation and stress in the mesolayers, which are unknown at this stage. 
Note that the matrices in (\ref{PBS_MesoFields}) are consistent with the volume average identities 
\begin{equation}
	\frac{1}{2} \left( \obfF^+ +  \obfF^- \right) = \obfF, \quad\quad 
	\frac{1}{2} \left( \obfP^+ +  \obfP^- \right) = \obfP,
	\label{PBS_AVEIDS}
\end{equation}
and the deformation- and traction-continuity conditions
\begin{equation}
	\obfF^+ -  \obfF^- = 2 \ogamma^* \bfe_1 \otimes \bfe_2, \quad\quad
	\obfP^+ \bfe_2 = \obfP^- \bfe_2 = \obfP \bfe_2.
	\label{PBS_PBConds}
\end{equation}
It is recalled that the tensors $\obfF$ and $\obfP$ in $(\ref{PBS_AVEIDS})$ and $(\ref{PBS_PBConds})$ are given by $(\ref{AL_MacroFields})_1$ and $(\ref{AL_MacroFields})_2$, respectively, and that the vector $\bfe_2$ in $(\ref{PBS_PBConds})$ stands for the lamination direction of the mesolayers. 
In addition, given that $\det \obfF^+ = \det \obfF^- = \det \obfF$, it follows from $(\ref{MSevol})$ that the volume fraction and the orientation of the mesolayers remain unchanged throughout the deformation. In contrast, making use of $\bfN=\bfe_1$ and expressions $(\ref{PBS_MesoFields})_1$ for $\obfF^+$ and $\obfF^-$ in $(\ref{MSevol})$, it is found that the lamination directions $\bfn^+$ and $\bfn^-$ of the microlayers in the two different sets of mesolayers are given by 
\begin{equation}
	\bfn^+ = \frac{\left(\obfF^+ \right)^{-T} \bfe_1 }{ \Big{|}\left(\obfF^+\right)^{-T} \bfe_1 \Big{|} }, \quad\quad
	\bfn^- = \frac{\left(\obfF^-\right)^{-T} \bfe_1 }{ \Big{|}\left(\obfF^-\right)^{-T} \bfe_1 \Big{|} }.
	\label{PBS_MSevol}
\end{equation}
It follows from $(\ref{PBS_MSevol})$ that the microlayers in the two sets of mesolayers rotate about $\bfe_3$ by the same angle $\phi$, but in opposite directions (see Fig. \ref{Sch_AL}(c)), such that $\cos \phi = \bfn^+ \cdot \bfe_1 = \bfn^- \cdot \bfe_1$.

The unknown components of the matrices in $(\ref{PBS_MesoFields})$ are determined from the piece-wise uniform solution of subsection \ref{PUF} for the homogenized behavior of each type of mesolayers, coupled with the elastoplastic constitutive equations of the Appendix for the local material behavior of the underlying microlayers. 
Given, however, the symmetries of these matrices, is suffices to consider only one type of mesolayers. Thus, focusing on the mesolayers of the type +, and taking into account the corresponding deformation- and traction-continuity conditions at the phase boundaries of the microlayers, as well as the plane-strain character of the deformation, it follows that the deformation gradient and first Piola-Kirchhoff stress tensors in the microlayers of these mesolayers (of the type +) have the following matrix forms
\begin{equation}
	\left[ \oF^{(r)} \right] =
	\begin{bmatrix}
		\oF^{(r)}_{11} & \ogamma^* & 0 \\
		\oF^{(r)}_{21} & \ol_2 & 0 \\
		0 & 0 & 1
	\end{bmatrix}, \quad\quad
	\left[ \oP^{(r)} \right] =
	\begin{bmatrix}
		0 & \oP^{(r)}_{12} & 0 \\
		\oG^* & \oP^{(r)}_{22} & 0 \\
		0 & 0 & \oP^{(r)}_{33}
	\end{bmatrix}
	\label{PBS_MicroFields}
\end{equation}
with $r=1,2$. Thus, the volume average identities for these variables over the corresponding mesolayers reduce to
\begin{equation}
	c^{(1)}_0 \oF^{(1)}_{11} + c^{(2)}_0 \oF^{(2)}_{11} = \ol_1, \quad\quad 
	c^{(1)}_0 \oF^{(1)}_{21} + c^{(2)}_0 \oF^{(2)}_{21} = 0,
	\label{PBS_AVEF}
\end{equation}
and
\begin{equation}
	c^{(1)}_0 \oP^{(1)}_{12} + c^{(2)}_0 \oP^{(2)}_{12} = 0, \quad\quad 
	c^{(1)}_0 \oP^{(1)}_{22} + c^{(2)}_0 \oP^{(2)}_{22} = \oP_{2}, \quad\quad 
	c^{(1)}_0 \oP^{(1)}_{33} + c^{(2)}_0 \oP^{(2)}_{33} = \oP_{3}.
	\label{PBS_AVEP}
\end{equation}
Rewriting expressions $(\ref{PBS_AVEF})$ in the form
\begin{equation}
	\oF^{(1)}_{11} = \left( \ol_1 - c^{(2)}_0 \oF^{(2)}_{11} \right) / c^{(1)}_0, \quad\quad 
	\oF^{(1)}_{21} = - c^{(2)}_0 \oF^{(2)}_{21} / c^{(1)}_0,
	\label{PBS_VAF}
\end{equation}
and taking into account the fact that $\ol_2$ is treated here as the loading parameter, the unknown deformation-gradient components reduce to $\ol_1$, $\ogamma^*$, $\oF^{(2)}_{11}$, and $\oF^{(2)}_{21}$. 
These unknowns are determined in terms of $\ol_2$ by means of a system of ordinary differential equations of the following form
\begin{align}
	&\tilde{\cal{L}}_{1111} \dot{\ol}_1 + \tilde{\cal{L}}_{1122} \dot{\ol}_2 + \tilde{\cal{L}}_{1112} \dot{\ogamma}^* = 0, \label{PBS_ODE1} \\ 
	&\tilde{\cal{L}}_{1112} \dot{\ol}_1 + \tilde{\cal{L}}_{2212} \dot{\ol}_2 + \tilde{\cal{L}}_{1212} \dot{\ogamma}^* = 0, \label{PBS_ODE2} \\
	&\dot{\oF}^{(2)}_{11} = {\cal A}^{(2)}_{1111} \dot{\ol}_1 + {\cal A}^{(2)}_{1122} \dot{\ol}_2 + {\cal A}^{(2)}_{1112} \dot{\ogamma}^* \label{PBS_ODE3} \\
	&\dot{\oF}^{(2)}_{21} = {\cal A}^{(2)}_{2111} \dot{\ol}_1 + {\cal A}^{(2)}_{2122} \dot{\ol}_2 + {\cal A}^{(2)}_{2112} \dot{\ogamma}^*.
	\label{PBS_ODE4}
\end{align}
Equations $(\ref{PBS_ODE1})$ and $(\ref{PBS_ODE2})$ express the conditions $\dot{\oP}^{+}_{11}=0$ and $\dot{\oP}^{+}_{12}=0$, respectively, and follow from $(\ref{SFRF})_1$, whereas equations $(\ref{PBS_ODE3})$ and $(\ref{PBS_ODE4})$ are obtained from $(\ref{FBARrRF2})_2$. 
It is remarked that, with a little abuse of notation, the coefficients in $(\ref{PBS_ODE1})$ and $(\ref{PBS_ODE2})$ represent components of the incremental effective modulus tensor $\widetilde{\bmL}$ of the mesolayers of the type + and, accordingly, the coefficients in $(\ref{PBS_ODE3})$ and $(\ref{PBS_ODE4})$ represent components of the deformation-concentration tensor $\bmA^{(2)}$ in the microlayers of type 2 in these mesolayers. 
Recall that the tensors $\widetilde{\bmL}$ and $\bmA^{(2)}$ are given by expressions $(\ref{SFRF})_2$ and $(\ref{ArTens})_2$, respectively, in terms of the phase modulus tensors $\bmL^{(r)}$ of the elastoplastic microlayers, which are in turn obtained from expressions $(\ref{ArTens})_2$, evaluated at the corresponding deformation gradients $(\ref{PBS_MicroFields})$.
It is also remarked that the superimposed dot over quantities in equations $(\ref{PBS_ODE1}) - (\ref{PBS_ODE4})$ indicates the derivative of these quantities with respect to the loading parameter $\ol_2$. 
Note that, when evaluated along the principal equilibrium path of the previous subsection, the system of ODEs $(\ref{PBS_ODE1})-(\ref{PBS_ODE4})$ reduces to
\begin{equation}
	\tilde{\cal{L}}_{1111} \dot{\ol}_1 + \tilde{\cal{L}}_{1122} \dot{\ol}_2 = 0, \quad\quad 
	\tilde{\cal{L}}_{1212} \dot{\ogamma}^* = 0, \quad\quad
	\dot{\oF}^{(2)}_{11} = {\cal A}^{(2)}_{1111} \dot{\ol}_1 + {\cal A}^{(2)}_{1122} \dot{\ol}_2, \quad\quad
	\dot{\oF}^{(2)}_{21} = 0.
	\label{PS_ODEs}
\end{equation}
It is observed from $(\ref{PS_ODEs})_2$ that if $\tilde{\cal{L}}_{1212} \ne 0$ then $\dot{\ogamma}^* = 0$.
Hence, in the strongly elliptic regime of the laminate, within which $\tilde{\cal{L}}_{1212} > 0$, we have that $\ogamma^* = 0$ and, in extend, the solution $(\ref{PBS_MesoFields})$ under consideration here coincides with the principal solution $(\ref{AL_MacroFields})$ of the previous subsection. 
The first instance at which the system of ODEs $(\ref{PBS_ODE1})-(\ref{PBS_ODE4})$ admits a non-trivial solution for $\ogamma^*$ as the loading parameter $\ol_2$ increases and/or decreases from its original value $\ol_2=1$ in the undeformed state (Fig. \ref{Sch_AL}(a)) is when $\ol_2$ reaches a critical value, say $\ol^{cr}_2$, for which the macroscopic LOE condition $(\ref{AL_LOEcond})$ is first met. This value of $\ol_2$ defines the onset of a bifurcation from the principal to the post-bifurcation solution, which leads in turn to the formation of the mesoscopic lamellar domains of Fig. \ref{Sch_AL}(c).
Thus, the unknown deformation gradient components $\ol_1$, $\ogamma^*$, $\oF^{(2)}_{11}$, and $\oF^{(2)}_{21}$ are obtained in terms of $\ol_2$ by integrating the system of ODEs $(\ref{PBS_ODE1}) - (\ref{PBS_ODE2})$, together with the associated elastoplastic constitutive relations of the Appendix, along any given loading path, starting from the critical value $\ol_2 = \ol^{cr}_2$. 
The initial conditions for this integration, i.e., the values of these unknowns at $\ol_2 = \ol^{cr}_2$, are determined from the principal solution of the previous subsection. 
The unknown stress components in $(\ref{PBS_MicroFields})_2$ are obtained as a byproduct of this integration procedure (see subsection \ref{NI}), while their counterparts in $(\ref{PBS_MesoFields})_2$, including the macroscopic stress components $\oP_2$ and $\oP_3$, are determined by making use of the volume average identities $(\ref{PBS_AVEP})$.

\section{Results and Discussion}
\label{RD}

In this section, we make use of the constitutive relations developed in the preceding section in order to study the effect of fiber plasticity on the macroscopic response and domain formation in soft biological composites.

For simplicity, we focus our considerations on the class of laminates comprised of a purely elastic phase 1 and an elastoplastic phase 2, assuming that the elastic behavior of each phase $r$, with $r=1,2$, is described by means of the Neo-Hookean stored-energy function (\ref{WrNH}). In addition, we assume that plastic yielding in phase 2 is governed by the von Mises yield criterion $(\ref{YC})$, along with the following isotropic stain-hardening law
\begin{equation}\label{Eqn: specific hardening law}
	\Sigma_{y}^{(2)}= \Sigma_{0}^{(2)} + h^{(2)} \oeps^{(2,p)},
\end{equation}
where $\Sigma_{0}^{(2)}$ is the initial yield stress of phase 2, $h^{(2)}$ is the corresponding (constant) hardening modulus, and $\oeps^{(2,p)}$ is the accumulated plastic strain in phase 2. 
In order to model fiber plasticity, we focus our attention on the case that the elastoplastic phase (fibers) is elastically stiffer than the purely elastic phase (matrix), such that $\mu^{(2)} \ge \mu^{(1)}$ and $\kappa^{(2)} \ge \kappa^{(1)}$. 
Given that the actual biological composites of interest are typically subjected to unidirectional, cyclic loading conditions along the direction of the fibers, in this section we also restrict our considerations to  aligned, plane-strain, non-monotonic loading conditions, such that the applied macroscopic stretch $\ol_2$ is first increased monotonically from its original value $\ol_2 = 1$ up to a maximum value $\ol^{max}_2$, and then decreased monotonically from $\ol_2=\ol^{max}_2$ down to a minimum value $\ol^{min}_2$. Results are presented in the sequel for various values of $\ol^{max}_2$, but the same value $\ol^{min}_2 = 0.5$ is used in all calculations. 
The results of this section are organized into two parts. In the first part, we focus our considerations on the principal and post-bifurcation behavior of the composite materials of interest, while in the second part, we present detailed results concerning the critical conditions for the onset of long-wavelength bifurcations, as determined by the macroscopic LOE condition $(\ref{LOECond})$ for the special case that $\ol^{(1,p)}_1 = \ol^{(1,p)}_2 \equiv  1$ and $\ol^{(1,e)}_1 \equiv \ol^{(1)}_1$.
For brevity, the principal and post-bifurcation solutions are labeled PS and PBS, respectively.
Whenever necessary, the critical conditions at which the laminate loses macroscopic ellipticity for the first time along the principal path are indicated by a superimposed dot.
Results corresponding to purely elastic and elastoplastic incremental deformations are represented by dashed and continuous lines, respectively.

\subsection{Principal and post-bifurcation behavior}
\label{MR}

\begin{figure}
	\begin{tabular}{cc}
		\centering
		\includegraphics[width=3.in]{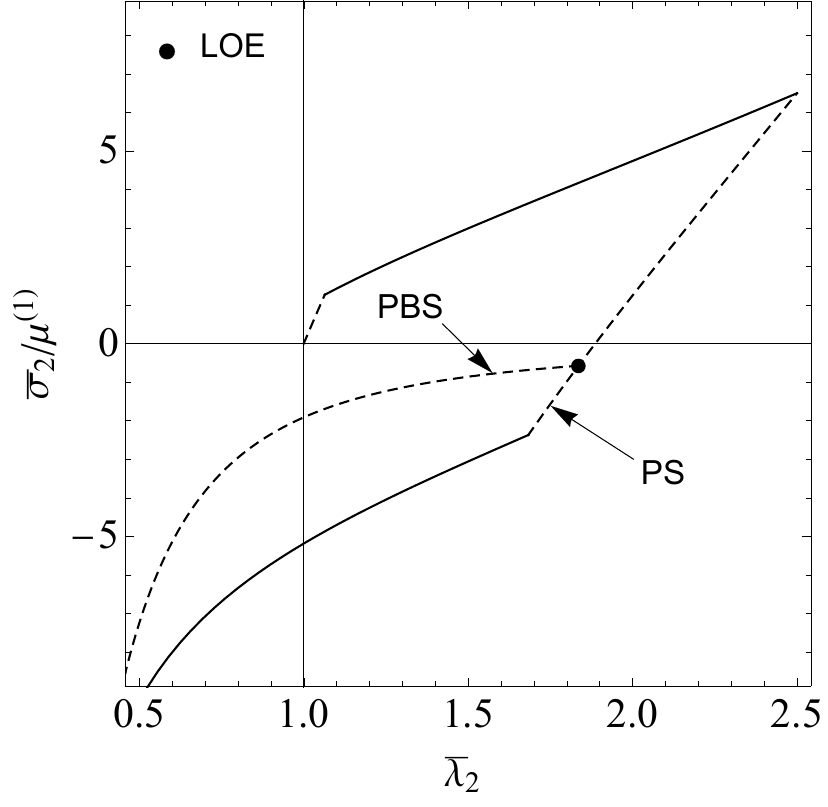} 
		& \includegraphics[width=3.in]{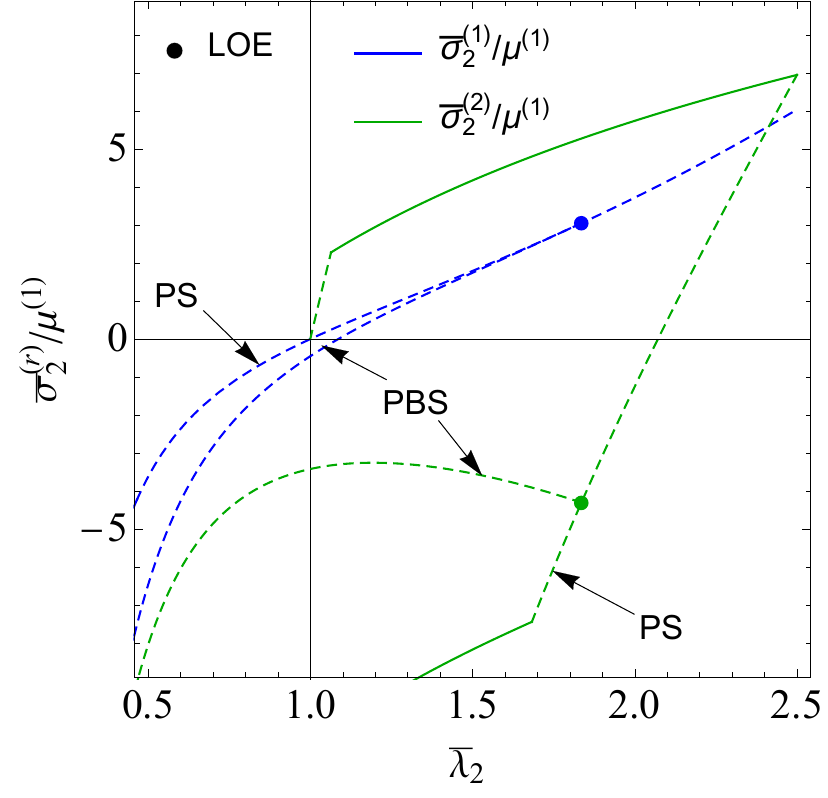} \\
		(a) & (b) \\
		\includegraphics[width=3.in]{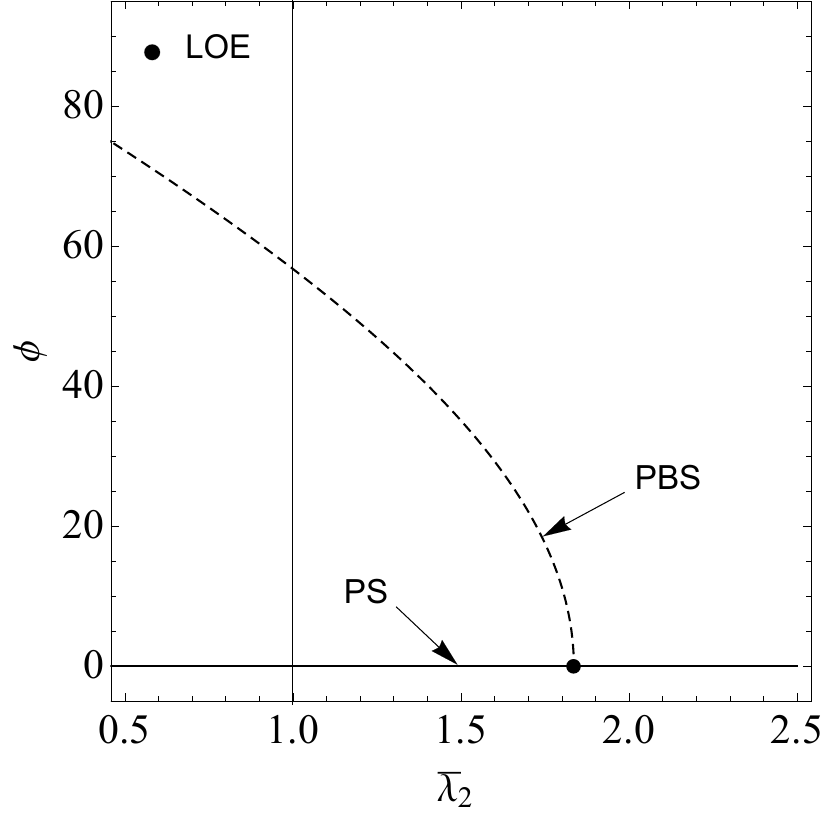} 
		& \includegraphics[width=3.in]{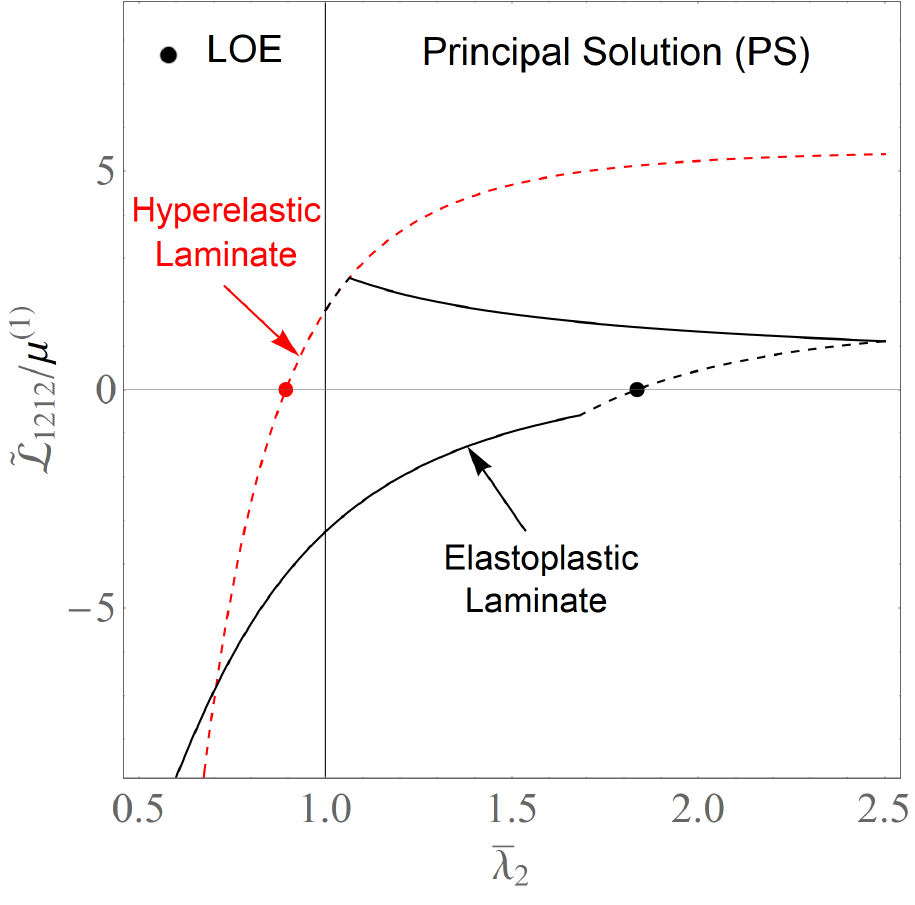} \\
		(c) & (d) \\
	\end{tabular}
	\caption{The response of a laminate comprised of elastic (phase 1) and elastoplastic (phase 2) layers with initial volume fractions $c_0^{(1)}=c_0^{(2)}=0.5$ and material properties $\mu^{(2)} = 10 \mu^{(1)}$,  $\kappa^{(1)} = \kappa^{(2)} = 100 \mu^{(1)}$, $\Sigma_0^{(2)} = \mu^{(2)}/5$, and $h^{(2)}=\mu^{(2)}/2$, under aligned plane-strain tension up to a maximum stretch $\ol^{max}_2 = 2.5$, followed by tensile unloading and subsequent compressive loading down to a minimum stretch $\ol^{min}_2 = 0.5$. 
		The principal and post-bifurcation solutions, labeled PS and PBS, respectively, are plotted versus $\ol_2$ for (a) the normalized macroscopic Cauchy stress $\osig_{2} / \mu^{(1)}$, (b) its local counterparts $\osig_{2}^{(1)} / \mu^{(1)}$ and $\osig_{2}^{(2)} / \mu^{(1)}$, 
		and (c) the lamination-orientation angle $\phi$ (in degrees) of the
		microlayers (see Fig. \ref{Sch_AL}(c)). (d) The principal solution (PS) for the normalized effective modulus $\tilde{\cal{L}}_{1212} / \mu^{(1)}$ of the elastoplastic laminate is plotted (black curve) versus $\ol_2$ and compared with its counterpart for the corresponding laminate with purely elastic phases (red curve).
		The parts of curves corresponding to purely elastic incremental deformations are drawn dashed, while those involving plastic yielding are shown continuous.
		The superimposed dots indicate the critical conditions for macroscopic loss of ellipticity (LOE).}
	\label{MR_h5}
\end{figure}

Figure \ref{MR_h5} shows results for an elastoplastic laminate with phase volume fractions $c_0^{(1)} = c_0^{(2)}=0.5$ and local material properties $\mu^{(2)} = 10 \mu^{(1)}$,  $\kappa^{(1)} = \kappa^{(2)} = 100 \mu^{(1)}$, $\Sigma_0^{(2)} = \mu^{(2)}/5$, and $h^{(2)}=\mu^{(2)}/2$, 
subjected initially to a monotonically increasing stretch $\ol_2$ up to its maximum value $\ol^{max}_2 = 2.5$ and subsequently to a monotonically decreasing $\ol_2$ down to its minimum value  $\ol^{min}_2 = 0.5$.
Figs. \ref{MR_h5}(a) and (b) show plots of the macroscopic Cauchy stress component $\osig_{2}$ and its local counterparts $\osig^{(1)}_{2}$ and $\osig^{(2)}_{2}$, normalized by $\mu^{(1)}$, versus the applied stretch $\ol_2$. 
It is remarked in the context of these plots that, during the tensile loading stage, i.e., as $\ol_2$ increases from its original value $\ol_2 = 1$, the deformations in the elastoplastic phase 2 are purely elastic up to the point where the slopes of the curves for $\osig_{2}$ and $\osig^{(2)}_{2}$ exhibit a sudden drop for the first time, indicating the onset of plastic yielding in this phase.
From this point and up to the maximum applied stretch $\ol^{max}_2 = 2.5$, phase 2 undergoes continuous plastic deformations. 
It should also be noticed from Fig. \ref{MR_h5}(b) that the stress $\osig^{(1)}_{2}$ in phase 1 remains unaffected by plastic yielding in phase 2, which is consistent with the corresponding remark made earlier in subsection \ref{PSol} that, for the given loading conditions, the principal solution for the stress-strain response of each constituent phase is independent from that of the other. 
As the applied stretch $\ol_2$ decreases monotonically from its maximum value $\ol^{max}_2 = 2.5$, shortly after the initiation of macroscopic compression ($\osig_2 < 0$) and while the incremental deformation of the composite is still purely elastic, a critical state is reached at which the laminate loses macroscopic ellipticity and the associated principal equilibrium solution reaches a bifurcation point, indicated by the superimposed dots in the plots of Fig. \ref{MR_h5}.
Quite remarkably, as can be seen from these results, the macroscopic critical stretch $\ol^{cr}_2$ at this state is extensive, i.e., $\ol^{cr}_2 \approx 1.85 > 1$.
In addition, it is observed from Figs. \ref{MR_h5}(a) and (b) that the critical macroscopic stress $\osig_{2}^{cr}$ and the corresponding stress $\osig^{(2, cr)}_{2}$ in the elastoplastic phase are both compressive, whereas the associated stress $\osig^{(1,cr)}_{2}$ in the elastic phase is tensile. 
As can be seen from the results of Fig. \ref{MR_h5}(a), the PBS for the macroscopic response of the laminate is much softer than the corresponding PS, especially right after the bifurcation point. 
This difference may be easily understood by bearing in mind that the PBS involves the formation of mesoscopic lamellar domains (Fig. \ref{Sch_AL}(c)), whereas the PS implies that the layers remain aligned with the loading direction (Fig. \ref{Sch_AL}(b)) at all time.
The formation and evolution of domains in the context of the PBS acts as a stress-relaxation mechanism which allows the macroscopic deformation to be accommodated primarily through layer rotation combined with intense shear deformation of the soft phase, parallel to the layers. By contrast, the PS involves necessarily equal contraction of the stiff and soft layers along the loading direction, which requires the application of a much greater axial stress. 
The PBS for the evolution of the lamination orientation angle $\phi$ (see Fig. \ref{Sch_AL}(c)) of the microlayers is shown in Fig. \ref{MR_h5}(c). It is observed from these results that, at the onset of bifurcation, the microlayers undergo a rapid rotation away from the loading direction $\bfe_2$. 
The effect of this rotation is most dramatically manifested in the results of Fig. \ref{MR_h5}(b) for $\osig_{2}^{(2)}$, which show that the stiff microlayers undergo unloading right after the onset of bifurcation. 
It is important to emphasize, however, that no unloading is observed in the macroscopic post-bifurcation behavior of the composite, as may be inferred from the slope of the PBS curve for $\osig_{2}$ in Fig. \ref{MR_h5}(a), which is negative at the onset of bifurcation and decreases continuously thereafter. 
In other words, the macroscopic stress $\osig_{2}$, which is compressive at the onset of bifurcation, keeps on decreasing monotonically with decreasing stretch $\ol_2$.
In order to gain further insight on the results of Figs. \ref{MR_h5}(a)-(c), and especially on the role of fiber plasticity on the macroscopic LOE, the PS for the normalized effective modulus $\tilde{\cal{L}}_{1212} / \mu^{(1)}$ of the elastoplastic laminate of this figure is plotted versus the applied stretch $\ol_2$ (black curve) in Fig. \ref{MR_h5}(d) and compared with that of the corresponding laminate with purely hyperelastic phases (red curve).
In the context of these results, it should be noticed that the modulus $\tilde{\cal{L}}_{1212}$ of the hyperelastic laminate increases (decreases) monotonically with increasing (decreasing) the applied stretch $\ol_2$, indicating that local elastic straining has a hardening (softening) effect on $\tilde{\cal{L}}_{1212}$ under increasing (decreasing) loading. 
It should also be noticed that this effect becomes progressively weaker (stronger) with increasing (decreasing) $\ol_2$, and is practically negligible for sufficiently large values of $\ol_2$, as may be inferred from the fact that the slope of $\tilde{\cal{L}}_{1212}$ increases asymptotically towards zero with increasing $\ol_2$. 
Thus, given that $\tilde{\cal{L}}_{1212}$ is strictly positive in the ground state ($\ol_2 = 1$) and that it decreases continuously with decreasing $\ol_2$, the hyperelastic laminate loses macroscopic ellipticity ($\tilde{\cal{L}}_{1212}=0$) at a contractile stretch ($\ol^{cr}_2 \approx 0.9 < 1$). 
It should be remarked that this behavior is quite typical for hyperelastic laminates with strongly elliptic constituents (see, e.g., \cite{TM85}, \cite{LPPC2009}).
From the dashed parts of the black curve of Fig. \ref{MR_h5}(d), it is similarly observed that local elastic straining has a hardening (softening) effect on the modulus $\tilde{\cal{L}}_{1212}$ of the elastoplastic laminate under increasing (decreasing) loading within its purely elastic regimes.
During the tensile loading stage, the two curves of Fig. \ref{MR_h5}(d) coincide and increase continuously with increasing loading up to the onset of plastic yielding in the elastoplastic composite, as expected. 
However, once plasticity sets in and up to the maximum applied stretch, the modulus $\tilde{\cal{L}}_{1212}$ of the elastoplastic laminate undergoes a continuous decrease, indicating that local plastic straining, which coexists with local elastic straining in this elastoplastic regime, has a dominant softening effect on $\tilde{\cal{L}}_{1212}$ with increasing macroscopic stretch $\ol_2$. 
This behavior leads to a value of $\tilde{\cal{L}}_{1212}$ at the maximum applied stretch $\ol^{max}_2 = 2.5$ which is considerably lower for the latter composite than for the former one. 
Upon unloading and subsequent compressive loading, the modulus $\tilde{\cal{L}}_{1212}$ of the elastoplastic laminate decreases monotonically due to the softening effect induced by local elastic straining and vanishes at an extensive stretch ($\ol^{cr}_2 \approx 1.85 > 1$).
For latter reference, it is also observed from the results of Fig. \ref{MR_h5}(d) that, under continued compression beyond the loss of macroscopic strong ellipticity, the incremental modulus $\tilde{\cal{L}}_{1212}$ of the elastoplastic laminate becomes negative and, after the re-initiation of plastic yielding (continuous, black curve), it keeps on decreasing, but at a noticeably slower rate than the decreasing rate of the corresponding modulus $\tilde{\cal{L}}_{1212}$ of the hyperelastic laminate (dashed, red curve).
This latter observation suggests that local plastic straining has a hardening effect on $\tilde{\cal{L}}_{1212}$ under decreasing loading conditions.
\begin{figure}
	\begin{tabular}{cc}
		\centering
		\includegraphics[width=3.in]{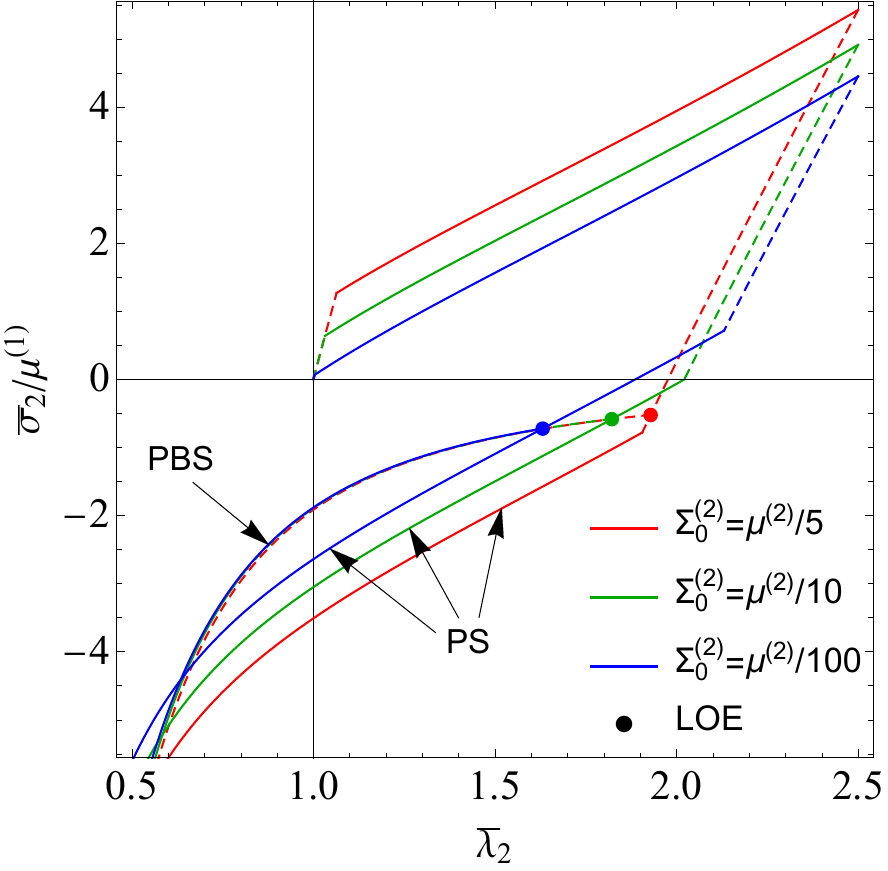} 
		& \includegraphics[width=3.in]{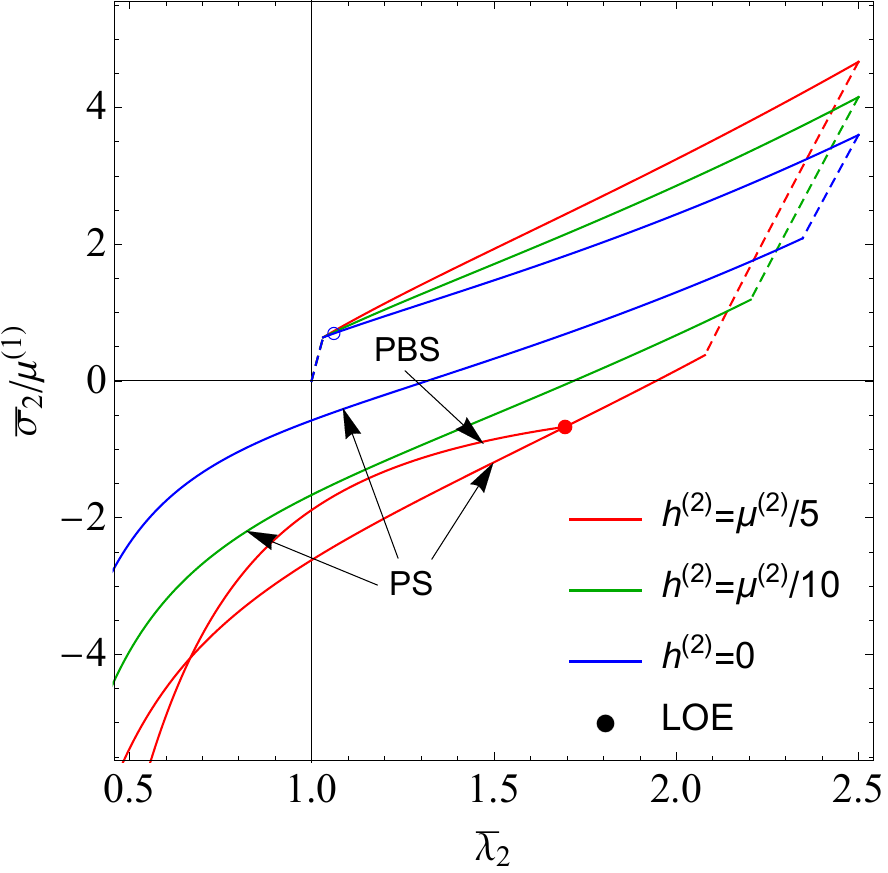} \\
		(a) & (b) \\
		\includegraphics[width=3.in]{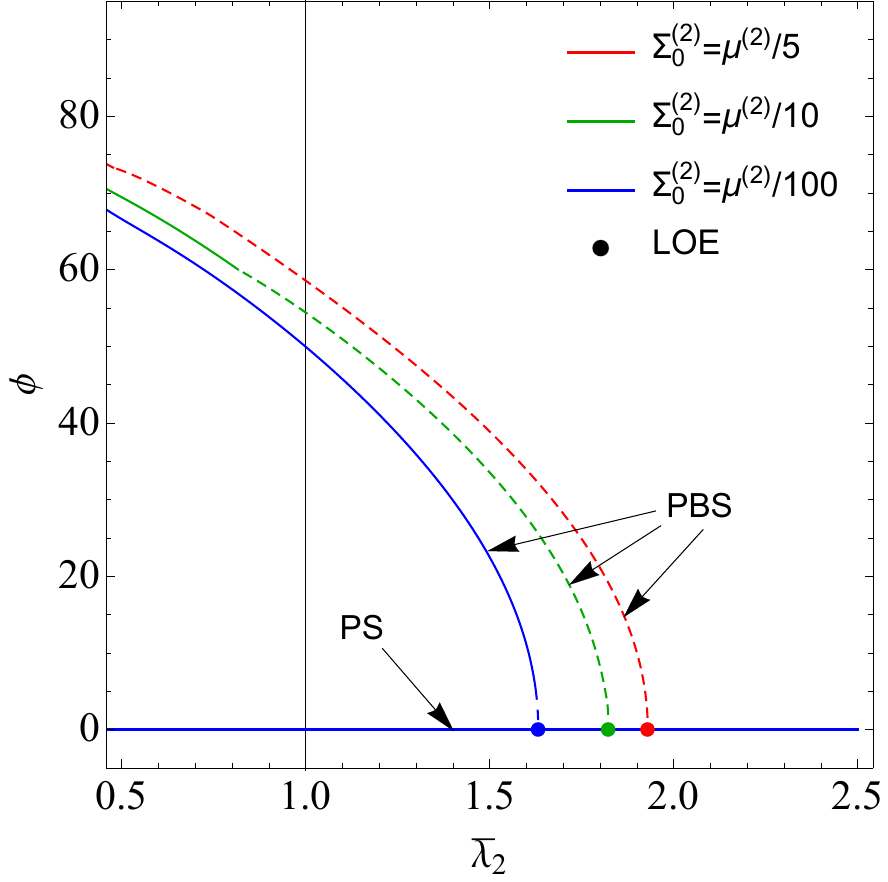} 
		& \includegraphics[width=3.in]{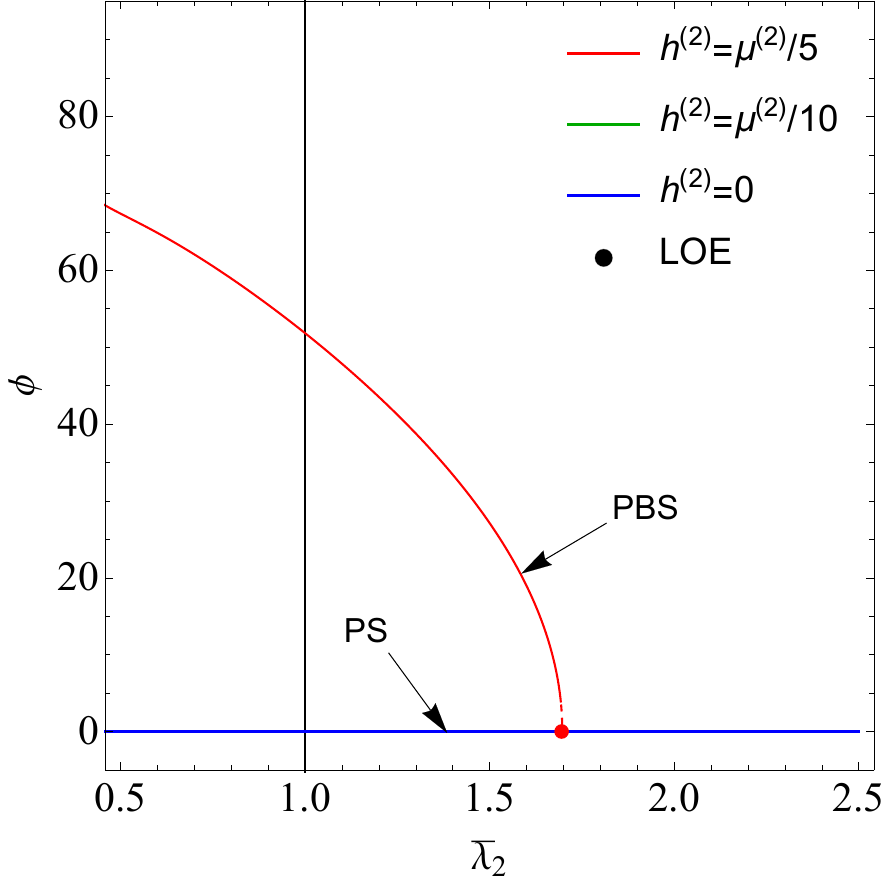} \\
		(c) & (d) \\
	\end{tabular}
	\caption{The macroscopic response of laminates with (a, c, e) the same local hardening rate  $h^{(2)}=\mu^{(2)}/4$ and different initial yield stresses $\Sigma^{(2)}_0 = \mu^{(2)}/5$, $\Sigma^{(2)}_0 = \mu^{(2)}/10$, and $\Sigma^{(2)}_0 = \mu^{(2)}/100$, and (b, d, f) the same initial yield stress $\Sigma^{(2)}_0 = \mu^{(2)}/10$ and different hardening rates $h^{(2)}=\mu^{(2)}/5$, $h^{(2)}=\mu^{(2)}/10$, and $h^{(2)}=0$ (perfect plasticity), under the same non-monotonic loading conditions as those considered in Fig. \ref{MR_h5}. The same initial phase volume fractions $c_0^{(1)} = c_0^{(2)}=0.5$ and local elastic properties $\mu^{(2)} = 10 \mu^{(1)}$ and $\kappa^{(1)} = \kappa^{(2)} = 100 \mu^{(1)}$ have been used in all cases. The principal and post-bifurcation solutions (PS and PBS, respectively) are shown versus the applied stretch $\ol_2$ for (a, b) the normalized macroscopic stress $\osig_{2} / \mu^{(1)}$ and (c, d) the lamination-orientation angle $\phi$ (in degrees) of the mesolayers (see Fig. \ref{Sch_AL}(c)). ...
	}
	\label{MR_h1_h01_h0}
\end{figure}
\begin{figure}
	\begin{tabular}{cc}
		\centering
		\includegraphics[width=3.in]{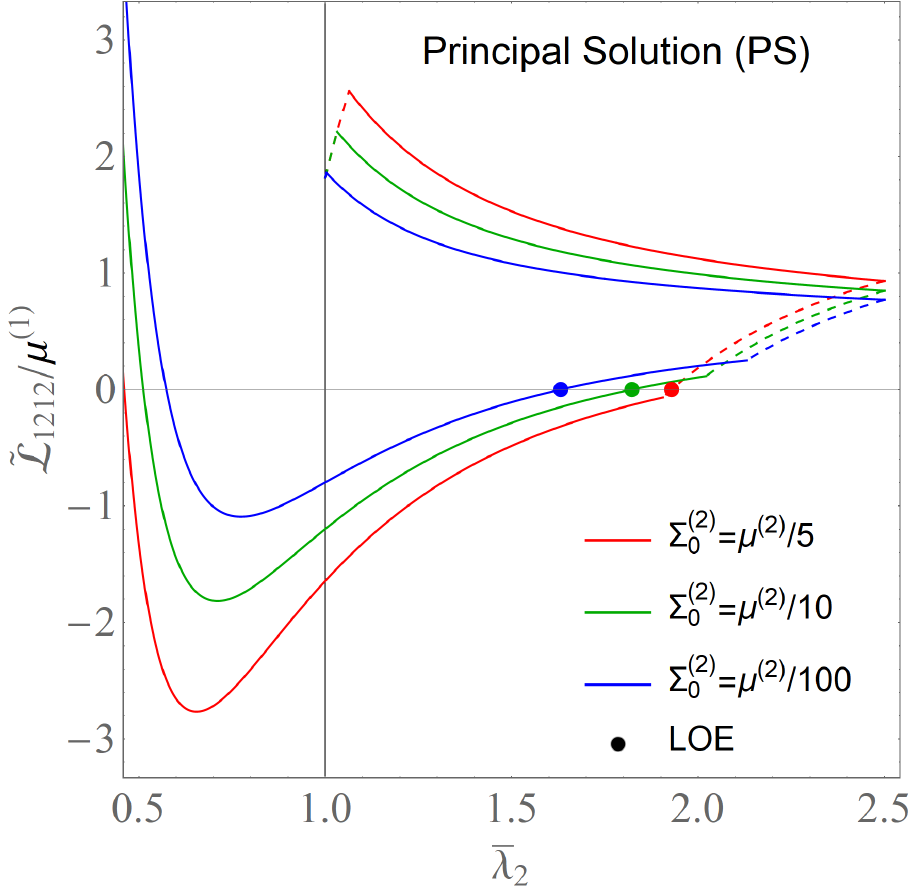} 
		& \includegraphics[width=3.in]{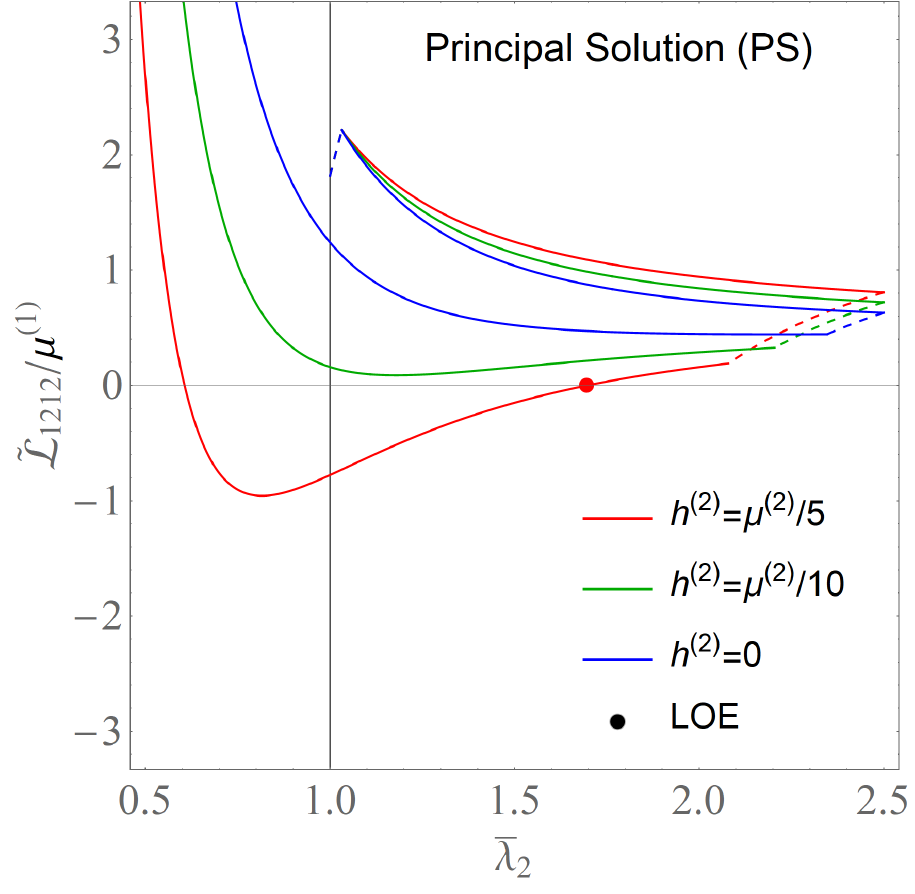} \\
		(e) & (f) \\
	\end{tabular}
	\caption*{Figure 6 (continued): ... (e, f) The principal solution (PS) is shown versus $\ol_2$ for the normalized incremental modulus $\tilde{\cal{L}}_{1212} / \mu^{(1)}$ of the laminate. Dashed and continuous lines are used in order to distinguish between purely elastic and elastoplastic regimes, respectively. 
		The superimposed dots indicate the critical conditions for macroscopic loss of ellipticity (LOE); the open cycle in part (b) indicates the conditions at which the elastoplastic phase loses ellipticity for the case $h^{(2)}=0$. 
	}
	\label{MR_h1_h01_h0_2}
\end{figure}

The effect of the local plastic properties on the macroscopic behavior of the elastoplastic laminates of interest is depicted in Fig. \ref{MR_h1_h01_h0}. 
This figure shows plots for the macroscopic response and microstructure evolution in composite materials with the same phase volume fractions $c_0^{(1)} = c_0^{(2)}=0.5$ and local elastic properties $\mu^{(2)} = 10 \mu^{(1)}$ and  $\kappa^{(1)} = \kappa^{(2)} = 100 \mu^{(1)}$, but different local plastic properties, under the same non-monotonic loading conditions considered previously in Fig. \ref{MR_h5}. 
More specifically, parts (a), (c), and (e) of this figure show results for laminates with the same local hardening modulus  $h^{(2)}=\mu^{(2)}/4$ and different initial yield stresses $\Sigma^{(2)}_0 = \mu^{(2)}/5$, $\Sigma^{(2)}_0 = \mu^{(2)}/10$, and $\Sigma^{(2)}_0 = \mu^{(2)}/100$, whereas parts (b), (d), and (f) show corresponding results for the same initial yield stress $\Sigma^{(2)}_0 = \mu^{(2)}/10$ and different hardening moduli $h^{(2)}=\mu^{(2)}/5$, $h^{(2)}=\mu^{(2)}/10$, and $h^{(2)}=0$ (perfect plasticity). 
The PS and PBS for the normalized macroscopic stress $\osig_{2} / \mu^{(1)}$ and lamination orientation angle $\phi$ for each one of these laminates are plotted in parts (a,b) and (c,d), respectively, versus the applied stretch $\ol_2$; corresponding plots of the PS for the normalized incremental modulus $\tilde{\cal{L}}_{1212} / \mu^{(1)}$ are presented in parts (e) and (f) of this figure. 
A general observation that can be made from the results of Figs. \ref{MR_h1_h01_h0}(a) and (b) is that the PS for the macroscopic stress-stretch response of the composite becomes softer for lower values of either $\Sigma^{(2)}_0$ or $h^{(2)}$, as expected. 
It is important to also observe, however, that the composite becomes less prone to lose macroscopic ellipticity for lower values of either $\Sigma^{(2)}_0$ or $h^{(2)}$.
More specifically, it is observed from Fig. \ref{MR_h1_h01_h0}(a) that the laminate loses macroscopic ellipticity at smaller stretches and stresses for lower values of $\Sigma^{(2)}_0$. 
In addition, it is observed from Fig. \ref{MR_h1_h01_h0}(b) that the composite loses macroscopic ellipticity for the case $h^{(2)}=\mu^{(2)}/5$, but it remains strongly elliptic for $h^{(2)}=\mu^{(2)}/10$ and $h^{(2)}=0$ (perfect plasticity). 
It should be remarked, however, that for the case $h^{(2)}=0$ the laminate loses strong ellipticity locally, in the elastoplastic phase, shortly after the onset of plastic yielding during tension, as indicated by the open circle in Fig. \ref{MR_h1_h01_h0}(c).
%
Additional calculations that have been carried out for the purpose of this work, but are not included here for brevity, show that LOE in the elastoplastic phase during tension takes place also for $h^{(2)} > 0$, but only if $h^{(2)}$ is sufficiently small.
The potential connection of this behavior with fibril micro-necking in tendons (Fig. \ref{Tendon}(c)) is worth investigating, but it lies beyond the scope of the present work. 
Another observation that should be made from Figs. \ref{MR_h1_h01_h0}(a) and (b) is that macroscopic LOE takes place at a compressive critical stress ($\osig_2^{cr} < 0$) for each of the cases that it does, despite the fact that the corresponding critical stretch is extensive ($\ol_2^{cr} < 0$). 
This observation suggests more generally that, just like hyperelastic laminates, elastoplastic laminates with strongly elliptic constituents do not lose macroscopic ellipticity under tensile loading or unloading. This point is further investigated in the context of the imperfection analysis presented in a forthcoming article (Part II).
The post-bifurcation curves of Figs. \ref{MR_h1_h01_h0}(a)-(d) exhibit qualitatively similar behavior to their counterpart in Figs. \ref{MR_h5}(a) and \ref{MR_h5}(c). It is interesting to observe, however, that the PBS curves for the three cases considered in Fig. \ref{MR_h1_h01_h0}(a) are practically indistibguishable from each other (within their common domain of existence). In addition, these curves are very similar to the corresponding PBS curve for $h^{(2)}=\mu^{(2)}/5$ in Fig. \ref{MR_h1_h01_h0}(b). This observation is consistent with the remark made earlier 
in the context of Fig. \ref{MR_h5} 
that the macroscopic deformation of the laminate is accommodated within the mesolayers (domains) primarily through microlayer rotation combined with intense longitudinal shear deformation in the soft microlayers (purely elastic phase), the properties of which are the same for all cases considered in Fig. \ref{MR_h1_h01_h0}. Thus, the results of this figure suggest that the effect of the plastic properties of the fiber phase (elastically stiffer layers) on the macroscopic, post-bifurcation response of the compose is rather negligible. However, as discussed in the previous paragraph and as elaborated further below, the effect of these properties on the critical conditions for the onset of these bifurcations is quite significant.
The results of Figs. \ref{MR_h1_h01_h0}(e) and (f) for the incremental effective modulus $\tilde{\cal{L}}_{1212}$ of the elastoplastic laminates under consideration here exhibit qualitatively similar behavior to the corresponding results for the elastoplastic laminate in Fig. \ref{MR_h5}(d). 
These results, as well as additional calculations that have been carried out for the purpose of this work, but are not presented here for brevity, show consistently that 
fiber plasticity has a dominant softening effect on $\tilde{\cal{L}}_{1212}$ under increasing loading and a hardening effect under decreasing loading conditions.
In addition, it is observed from Figs. \ref{MR_h1_h01_h0}(e) and (f) that both effects become stronger with decreasing either the initial yield stress $\Sigma^{(2)}_0$ (Fig. \ref{MR_h1_h01_h0}(e)) or the hardening modulus $h^{(2)}$ (Fig. \ref{MR_h1_h01_h0}(f)), which is consistent with the fact that lower values of either $\Sigma^{(2)}_0$ or $h^{(2)}$ lead to greater plastic and smaller elastic strains.
At the same time, however, as can be seen from Figs. \ref{MR_h1_h01_h0}(a) and (b), decreasing the  value of either $\Sigma^{(2)}_0$ or $h^{(2)}$ leads to a corresponding increase of the macroscopic stress $\osig_2$ at which plastic yielding begins during the reverse loading stage. 
This fact implies, in turn, that the strain range during reverse loading within which the softening effect due to elasticity can operate before the hardening effect induced by plasticity enters into play becomes smaller for lower values of either $\Sigma^{(2)}_0$ or $h^{(2)}$, as can be seen from Figs. \ref{MR_h1_h01_h0}(e) and (f). 
Note, in particular, that for the highest initial yield stress $\Sigma^{(2)}_0 = \mu^{(2)}/5$ in Fig. \ref{MR_h1_h01_h0}(e) the modulus $\tilde{\cal{L}}_{1212}$ vanishes within the purely elastic regime during compression, whereas for the smaller values $\Sigma^{(2)}_0 = \mu^{(2)}/10$ and $\Sigma^{(2)}_0 = \mu^{(2)}/100$ plastic yielding begins while the modulus $\tilde{\cal{L}}_{1212}$ is still positive and its rate of decrease towards zero is subsequently slowed down by the hardening effect of plastic straining with continued compression. 
The behavior of the modulus $\tilde{\cal{L}}_{1212}$ for the highest hardening rate $h^{(2)}=\mu^{(2)}/5$ in Fig. \ref{MR_h1_h01_h0}(f) is similar to that for the cases $\Sigma^{(2)}_0 = \mu^{(2)}/10$ and $\Sigma^{(2)}_0 = \mu^{(2)}/100$ in Fig. \ref{MR_h1_h01_h0}(e). Note, however, that for the smaller values of the hardening rate $h^{(2)}=\mu^{(2)}/10$ and $h^{(2)}=0$ in Fig. \ref{MR_h1_h01_h0}(f) the hardening effect induced by local plasticity on $\tilde{\cal{L}}_{1212}$ during compressive loading is so strong that the modulus $\tilde{\cal{L}}_{1212}$ remains positive throughout the loading and the laminate never loses macroscopic ellipticity.
\subsection{Critical conditions for the onset of bifurcations}
\label{LOME}

\begin{figure}
	\begin{tabular}{cc}
		\centering
		\includegraphics[width=3.in]{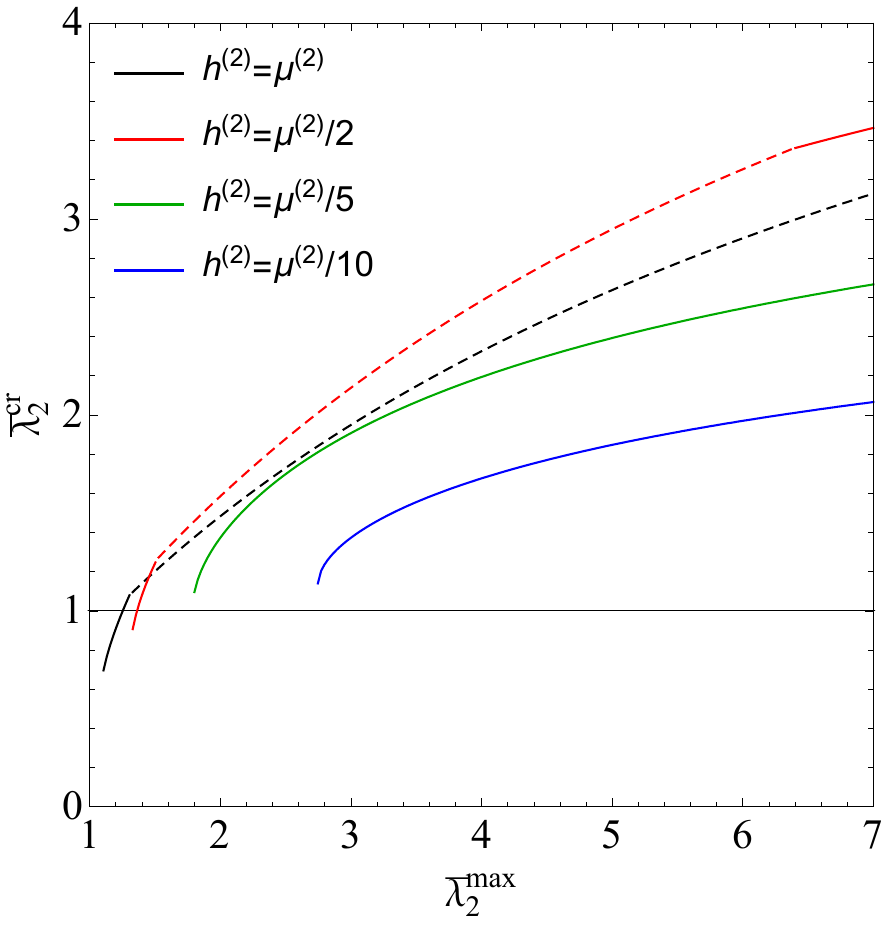} 
		& \includegraphics[width=3.25in]{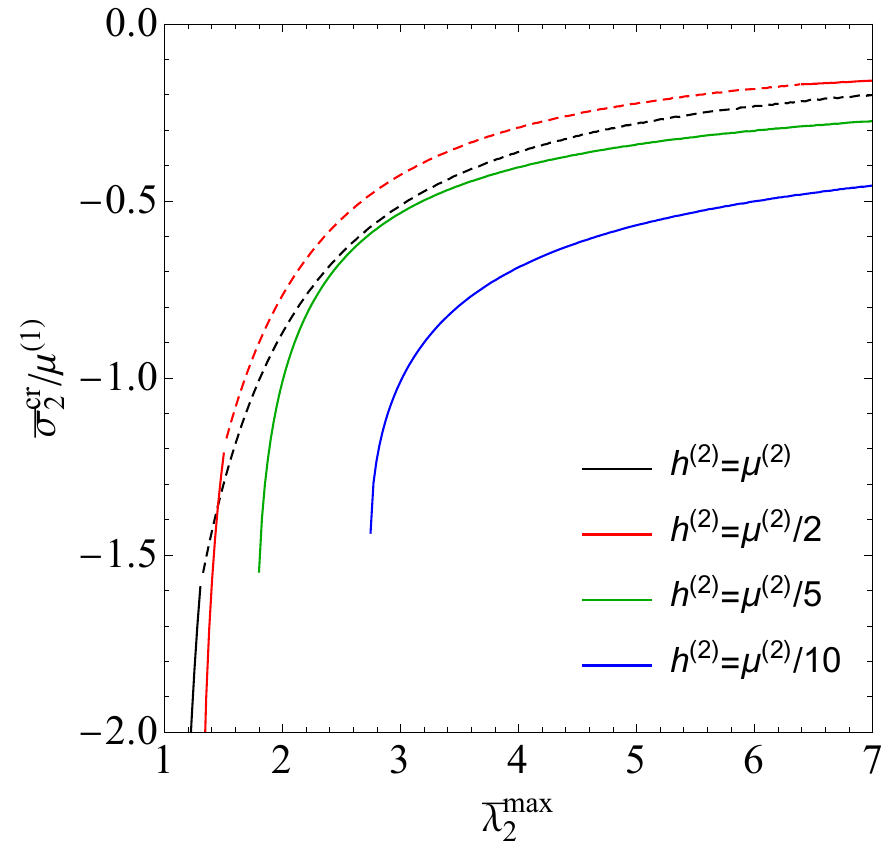} \\
		(a) & (b) \\
	\end{tabular}
	\caption{Results for the macroscopic LOE of elastoplastic laminates with local material properties $\mu^{(2)} = 10 \mu^{(1)}$,  $\kappa^{(1)} = \kappa^{(2)} = 100 \mu^{(1)}$, $\Sigma_0^{(2)} = \mu^{(2)}/10$, and initial phase volume fractions $c_0^{(1)}=c_0^{(2)}=0.5$, under aligned, plane-strain tension up to a maximum applied stretch $\ol^{max}_2$, followed by tensile unloading  and subsequent compressive loading. 
	(a) The macroscopic critical stretch $\ol^{cr}_{2}$ and (b) the corresponding normalized stress $\osig^{cr}_{2} / \mu^{(1)}$ are plotted as functions of $\ol^{max}_2$ for the cases $h^{(2)}=\mu^{(2)}$, $h^{(2)}=\mu^{(2)}/2$, $h^{(2)}=\mu^{(2)}/5$, and $h^{(2)}=\mu^{(2)}/10$. Critical conditions corresponding to macroscopic LOE in the purely elastic and elasplastic regimes are shown with dashed and continuous lines, respectively.
	}
	\label{LOECurves_vs_L2max_h2_10_5_2_1}
\end{figure}

In what follows, we focus our attention on the effects of the loading and material parameters on the critical conditions for the onset of bifurcations of the long-wavelength type in the elastoplastic laminates of interest, as captured by the loss of strong ellipticity of their homogenized behavior.
The results presented next, as well as the preceding results, with the exception of the case for $h^{(2)}=0$ in Figs. \ref{MR_h1_h01_h0}(b), (d), and (f), are all for laminates the constituent phases of which remain strongly elliptic (at least) up the onset of macroscopic LOE.
Figure \ref{LOECurves_vs_L2max_h2_10_5_2_1} shows plots of the macroscopic critical stretch $\ol^{cr}_{2}$ and normalized stress $\osig^{cr}_{2} / \mu^{(1)}$ as functions of the maximum applied stretch $\ol^{max}_2$ for elastoplastic laminates with different strain-hardening rates $h^{(2)}$. 
It is observed that every composite of this figure remains strongly elliptic for values of $\ol^{max}_2$ ranging from $\ol^{max}_2=1$ (monotonic compression) up to a certain value of $\ol^{max}_2$, depending on the specific value of $h^{(2)}$. 
This observation indicates that, for sufficiently small values of $\ol^{max}_2$, the hardening effect induced by local plasticity on the critical mode during the reverse loading stage is strong enough to overcome the corresponding softening effect due to local elasticity and, thus, prevent the vanishing of the incremental modulus $\tilde{\cal{L}}_{1212}$, as illustrated by the results of Fig. \ref{MR_h1_h01_h0}(f) for the special case that $h^{(2)} = \mu^{(2)} / 10$ and $\ol^{max}_2 = 2.5$. 
Note also that, the range of values of $\ol^{max}_2$ for which the laminate exhibits strongly elliptic behavior increases with decreasing the hardening rate $h^{(2)}$, indicating accordingly that the aforementioned hardening effect becomes stronger for smaller values of $h^{(2)}$, which is in turn consistent with the corresponding observation made earlier in the context of Fig. \ref{MR_h1_h01_h0}(f).
In the range of values of $\ol^{max}_2$ for which macroscopic LOE does take place, it is observed that the critical stretch $\ol^{cr}_{2}$ and stress $\osig^{cr}_{2}$ for each specific composite in Fig. \ref{LOECurves_vs_L2max_h2_10_5_2_1} increase monotonically and significantly with increasing $\ol^{max}_2$. 
Note, in particular, that the critical stress $\osig^{cr}_{2}$ in Fig. \ref{LOECurves_vs_L2max_h2_10_5_2_1}(b) becomes progressively closer to zero with increasing $\ol^{max}_2$ for each specific value of $h^{(2)}$, but it remains compressive ($\osig^{cr}_{2} < 0$) for all values of $h^{(2)}$.
The monotonic behavior of the curve in Fig. \ref{LOECurves_vs_L2max_h2_10_5_2_1} should be attributed primarily to the continuous softening of the modulus $\tilde{\cal{L}}_{1212}$ under increasing loading in the elastoplastic regime (Figs. \ref{MR_h5}(d), \ref{MR_h1_h01_h0}(e), and \ref{MR_h1_h01_h0}(f)), which leads to a value for $\tilde{\cal{L}}_{1212}$ at the end of the tensile loading stage that is closer to zero for higher values of $\ol^{max}_2$. 
Thus, it may be inferred from the results of this figure that the amount of plastic deformations produced during tension in the elastically stiffer layers (fibers), which increases with increasing $\ol^{max}_2$, has a significant effect on the critical conditions $\ol^{cr}_{2}$ and $\osig^{cr}_{2}$ for the onset of bifurcations in these composites. 
Another observation that should be made in the context of Fig. \ref{LOECurves_vs_L2max_h2_10_5_2_1} is that when LOE takes place under conditions of purely elastic straining (dashes parts of the curves), both the critical stretch $\ol^{cr}_2$ and stress $\osig^{cr}_2$ are higher for lower values of the hardening rate $h^{(2)}$, while when LOE occurs under conditions of elastoplastic straining (continuous parts of the curves), the opposite is true.

\begin{figure}
	\begin{tabular}{cc}
		\centering
		\includegraphics[width=3.1in]{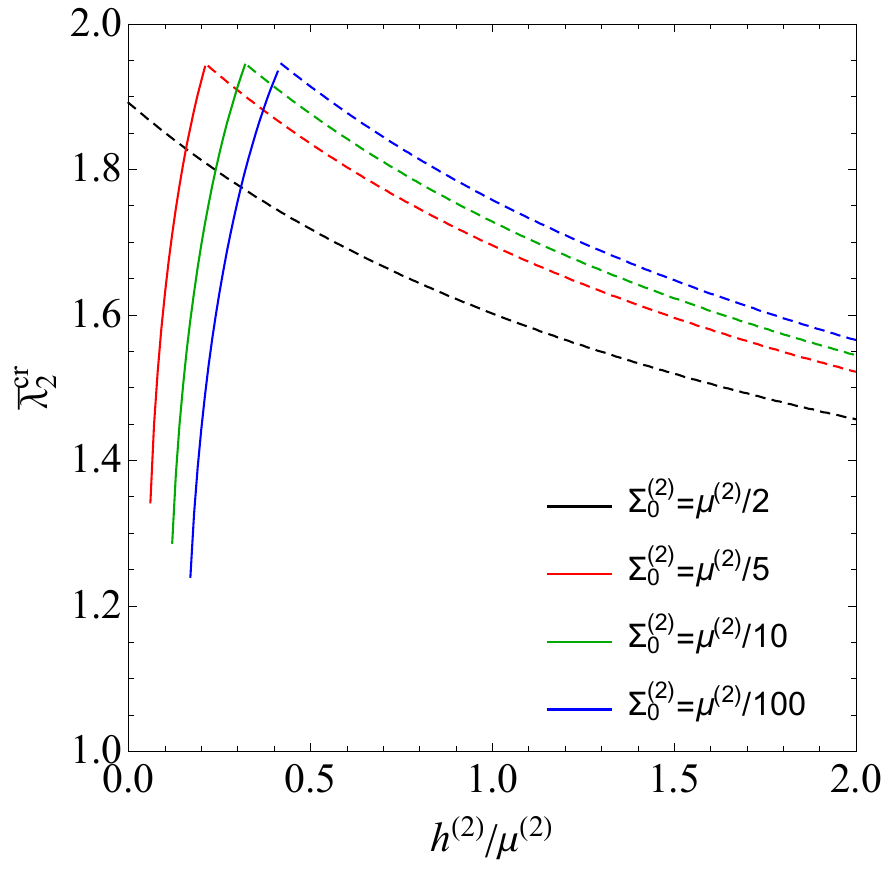} 
		& \includegraphics[width=3.2in]{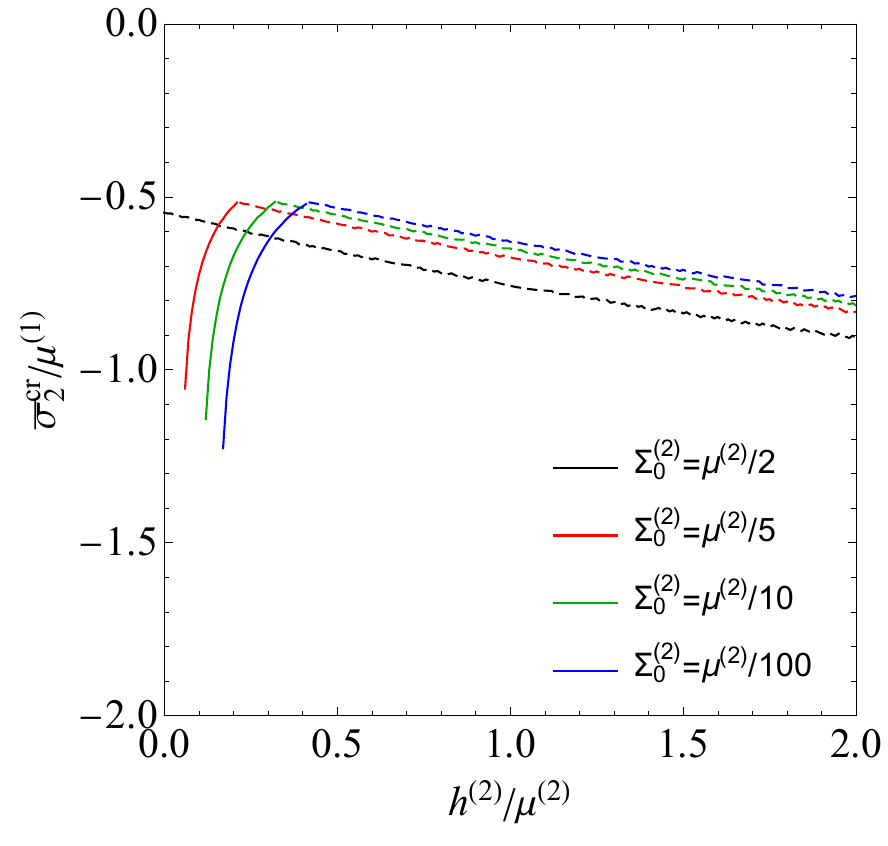} \\
		(a) & (b) \\
	\end{tabular}
	\caption{Macroscopic LOE curves for elastoplastic laminates with local material properties $\mu^{(2)} = 10 \mu^{(1)}$,  $\kappa^{(1)} = \kappa^{(2)} = 100 \mu^{(1)}$, $h^{(2)} = \mu^{(2)}/10$, and initial phase volume fractions $c_0^{(1)}=c_0^{(2)}=0.5$, under aligned, plane-strain tension up to a maximum applied stretch $\ol^{max}_2 = 2.5$, followed by tensile unloading and subsequent compressive loading. 
		(a) The macroscopic critical stretch $\ol^{cr}_{2}$ and (b) the corresponding normalized stress $\osig^{cr}_{2} / \mu^{(1)}$ are plotted as functions of the normalized hardening rate $h^{(2)} / \mu^{(2)}$ for the cases $\Sigma^{(2)}_0=\mu^{(2)} / 2$, $\Sigma^{(2)}_0=\mu^{(2)}/5$, $\Sigma^{(2)}_0=\mu^{(2)}/10$, and $\Sigma^{(2)}_0=\mu^{(2)}/100$. Critical conditions corresponding to macroscopic LOE in the purely elastic and elasplastic regimes are shown with dashed and continuous lines, respectively.
	}
	\label{LOECurves_vs_h2_S02_5_2_1_01}
\end{figure}

The influence of the strain-hardening rate $h^{(2)} $ on the critical conditions for macroscopic LOE is depicted more clearly in Fig. \ref{LOECurves_vs_h2_S02_5_2_1_01}, where $\ol^{cr}_{2}$ and $\osig^{cr}_{2} / \mu^{(1)}$ are plotted as functions of $h^{(2)} / \mu^{(2)}$ for laminates with different local yield stresses $\Sigma^{(2)}_0$ subjected to a maximum stretch $\ol^{max}_{2} = 2.5$.
Thus, it is observed from the results for each specific value of $\Sigma^{(2)}_0$ in this figure that the critical stretch $\ol^{cr}_{2}$ and stress $\osig^{cr}_{2}$ increase monotonically with decreasing $h^{(2)}$ when LOE takes place in the purely elastic regime (dashed parts of the curves) and decrease monotonically with decreasing $h^{(2)}$ when LOE occurs in the elastoplastic regime (continuous parts of the curves). 
These different trends in the two regimes reflect the opposite effects induced by fiber plasticity on the critical mode for LOE under increasing and decreasing loading conditions, namely, the softening effect under increasing and the hardening effect under decreasing loading, combined with the fact that both effects become stronger for lower values of $h^{(2)}$, as already discussed. 
In this connection, it should also be kept in mind that LOE in the elastic regime is influenced only by the former softening effect, whereas LOE in the elastoplastic regime is affected also by the latter hardening effect. 
Hence, it may be inferred from the results of this figure that the trend of the critical conditions for LOE in the elastic regime is determined by the former softening effect, while that for LOE in the elastoplastic regime is controlled by the latter hardening effect. 
It is also observed from Fig. \ref{LOECurves_vs_h2_S02_5_2_1_01} that the composites with the highest initial yield stress $\Sigma^{(2)}_0=\mu^{(2)} / 2$ lose macroscopic ellipticity under purely elastic incremental deformations for all values of $h^{(2)}$. 
On the other hand, the composites with lower values of $\Sigma^{(2)}_0$ remain strongly elliptic up to a certain value of $h^{(2)}$, depending on $\Sigma^{(2)}_0$, beyond which they lose macroscopic ellipticity continuously with increasing $h^{(2)}$, initially in the elastoplastic regime and then in the purely elastic regime, with the transition from the former to the latter type of LOE taking place at lower values of $h^{(2)}$ for higher values of $\Sigma^{(2)}_0$.

Another interesting observation that should be made from the results of Fig. \ref{LOECurves_vs_h2_S02_5_2_1_01} is that, for any given value of $h^{(2)}$ and increasing values of $\Sigma^{(2)}_0$, both the critical stretch $\ol^{cr}_{2}$ and stress $\osig^{cr}_{2}$ increase when LOE takes place in the elastoplastic regime and decrease when LOE occurs in the purely elastic regime. 
This observation, which has been confirmed by means of more detailed plots of $\ol^{cr}_{2}$ and $\osig^{cr}_{2} / \mu^{(1)}$ versus $\Sigma^{(2)}_0 / \mu^{(2)}$, which are omitted for brevity, suggests that the local plastic properties $\Sigma^{(2)}_0$ and $h^{(2)}$ have qualitatively similar effects on the macroscopic LOE of the composite. 
This conclusion may also be inferred from the analytical expression $(\ref{AL_LOEcond})_2$ for the effective incremental modulus $\tilde{\cal{L}}_{1212}$ of the laminate, the dependence of which on $\Sigma^{(2)}_0$ and $h^{(2)}$ enters implicitly through the influence of these parameters of the local plastic stretches $\ol_1^{(2,p)}$ and $\ol_2^{(2,p)}$.
Thus, it may be inferred from the preceding results that both the softening and the hardening effect induced by fiber plasticity on $\tilde{\cal{L}}_{1212}$ under conditions of increasing and decreasing loading, respectively, are stronger in composites with plastically more compliant fibers, i.e., for lower values of $\Sigma^{(2)}_0$ and $h^{(2)}$. 
In addition, it may be inferred from these results that macroscopic LOE tends to occur in the purely elastic regime for composites with sufficiently stiff fibers and in the elastoplastic regime for composites with moderately stiff fibers, whereas composites with sufficiently compliant fibers tend to remain strongly elliptic throughout the loading path.    
\begin{figure}
	\begin{tabular}{cc}
		\centering
		\includegraphics[width=3.in]{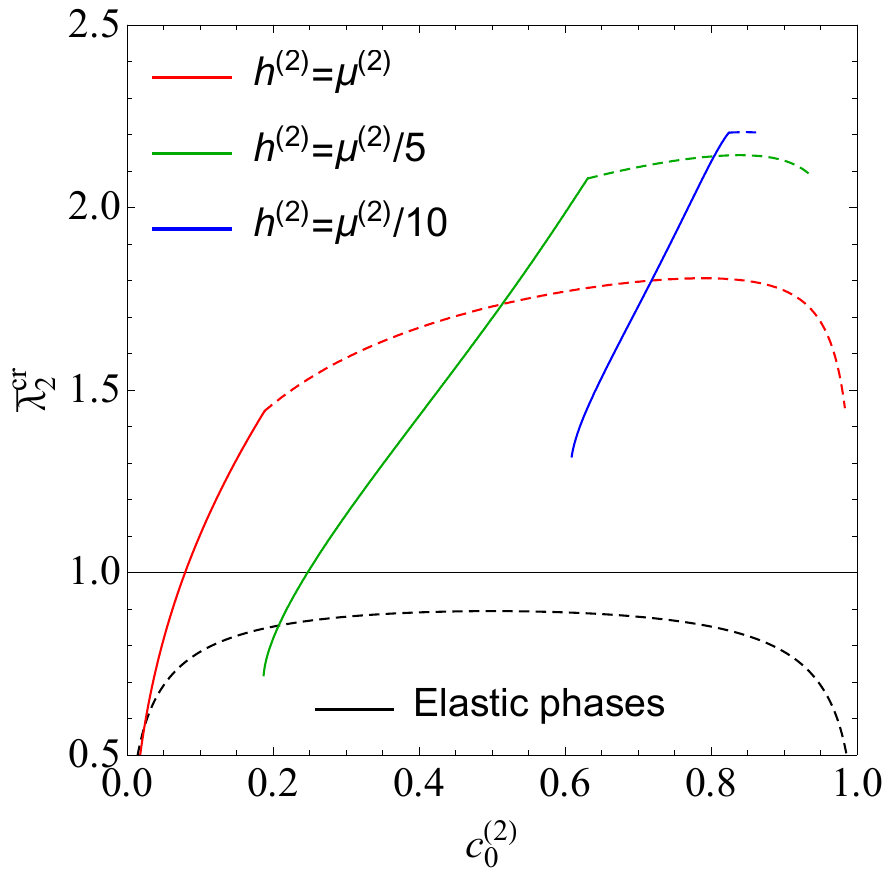} 
		& \includegraphics[width=3.in]{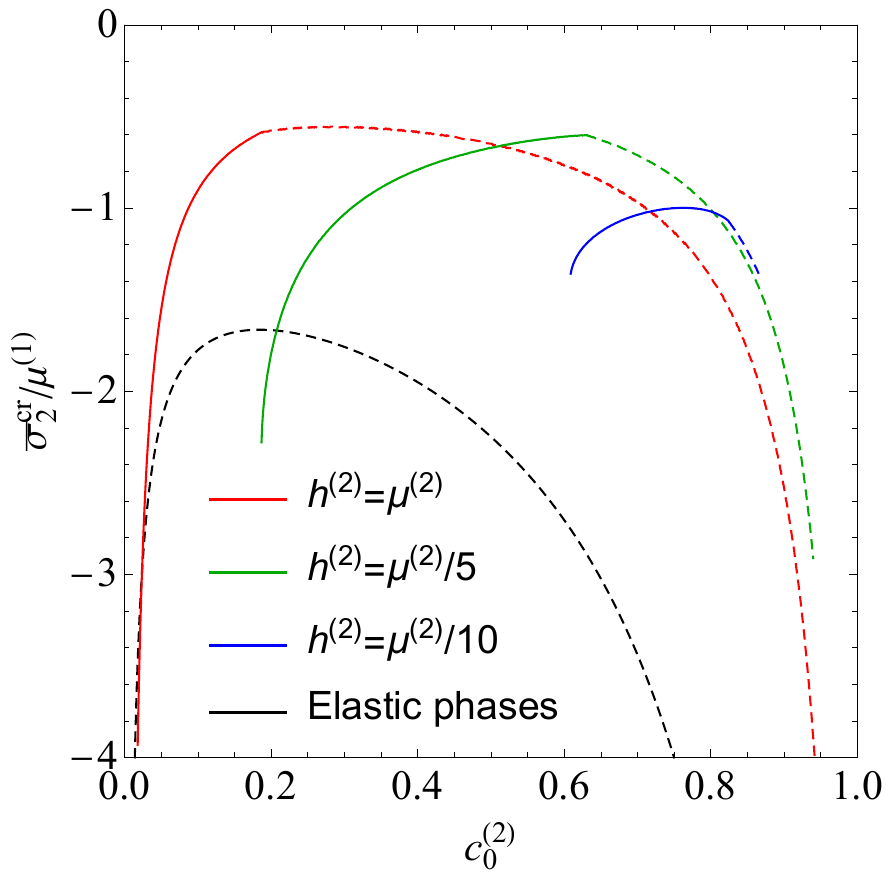} \\
		(a) & (b) \\
	\end{tabular}
	\caption{Macroscopic LOE curves for elastoplastic laminates with local material properties $\mu^{(2)} = 10 \mu^{(1)}$,  $\kappa^{(1)} = \kappa^{(2)} = 100 \mu^{(1)}$, and $\Sigma^{(2)}_0=\mu^{(2)}/10$, under aligned, plane-strain tension up to a maximum applied stretch $\ol^{max}_2 = 2.5$, followed by tensile unloading, and subsequent compressive loading. 
		(a) The macroscopic critical stretch $\ol^{cr}_{2}$ and (b) the corresponding normalized stress $\osig^{cr}_{2} / \mu^{(1)}$ are plotted as functions of the initial volume fraction $c_0^{(2)}$ of the elastoplastic phase for the cases $h^{(2)} = \mu^{(2)}$, $h^{(2)} = \mu^{(2)}/5$, and $h^{(2)} = \mu^{(2)}/10$. The critical curves for the corresponding laminates with purely elastic phases are also included in these plots for comparison. Critical conditions corresponding to macroscopic LOE in the purely elastic and elasplastic regimes are shown with dashed and continuous lines, respectively.
	}
	\label{LOECurves_vs_c02_h2_10_2_1}
\end{figure}

Figure \ref{LOECurves_vs_c02_h2_10_2_1} shows plots of the critical conditions $\ol^{cr}_{2}$ and $\osig^{cr}_{2} / \mu^{(1)}$ as functions of the volume fraction $c_0^{(2)}$ of the elastoplastic phase for laminates with different local hardening rates $h^{(2)}$ subjected to a fixed maximum stretch $\ol^{max}_2 = 2.5$.
For comparison, results are also shown for the corresponding laminate with purely elastic phases. 
It is observed from Fig. \ref{LOECurves_vs_c02_h2_10_2_1}(a) that the critical stretch $\ol^{cr}_{2}$ for the hyperelastic laminate is symmetric with respect to the value $c_0^{(2)} =0.5$ and decreases monotonically with either increasing or decreasing the volume fraction $c_0^{(2)}$ from this value. This observation suggests accordingly that the softening effect induced on the critical mode by local elastic straining under decreasing loading conditions becomes weaker with either increasing or decreasing $c_0^{(2)}$ from $c_0^{(2)} =0.5$. Note also that the critical stretch for the hyperelastic laminate tends to zero in the limits as $c_0^{(2)} \rightarrow 0$ and $c_0^{(2)} \rightarrow 1$, which is consistent with the fact that in these limits the composite reduces to the homogeneous and strongly elliptic phase 1 and 2, respectively.
On the other hand, the elastoplastic laminates of this figure do not lose macroscopic ellipticity up to a certain value of $c_0^{(2)} > 0$, as well as beyond a certain value of $c_0^{(2)} < 1$, both of which depend on the specific value of the hardening rate $h^{(2)}$. 
This observation indicates that, during the reverse loading stage, the hardening effect induced on the critical mode by plastic straining dominates over the corresponding softening effect due to elastic straining, which, as observed earlier, becomes weaker as the limiting values $c_0^{(2)} =0$ and $c_0^{(2)} =1$ are approached from their right and left, respectively.
The fact that the range of values of $c_0^{(2)}$ for which these elastoplastic laminates are strongly elliptic increases with decreasing the value of $h^{(2)}$ reflects once again the fact that the former hardening effect becomes stronger for smaller values of $h^{(2)}$. 
It is also interesting to observe from Fig. \ref{LOECurves_vs_c02_h2_10_2_1}(a) that, in contrast with the LOE curve for the hyperelastic laminate, which is symmetric about $c_0^{(2)} =0.5$, the LOE curves for the elastoplastic composites are skewed towards values of $c_0^{(2)}$ greater than $c_0^{(2)} =0.5$ and attain their maximum at values of $c_0^{(2)}$ that are closer to $c_0^{(2)} =1$. 
This observation suggests that fiber plasticity has a stronger effect on the macroscopic LOE of the composite for higher concentrations of the elastoplastic phase.
\begin{figure}[t]
	\begin{tabular}{cc}
		\centering
		\includegraphics[width=3.in]{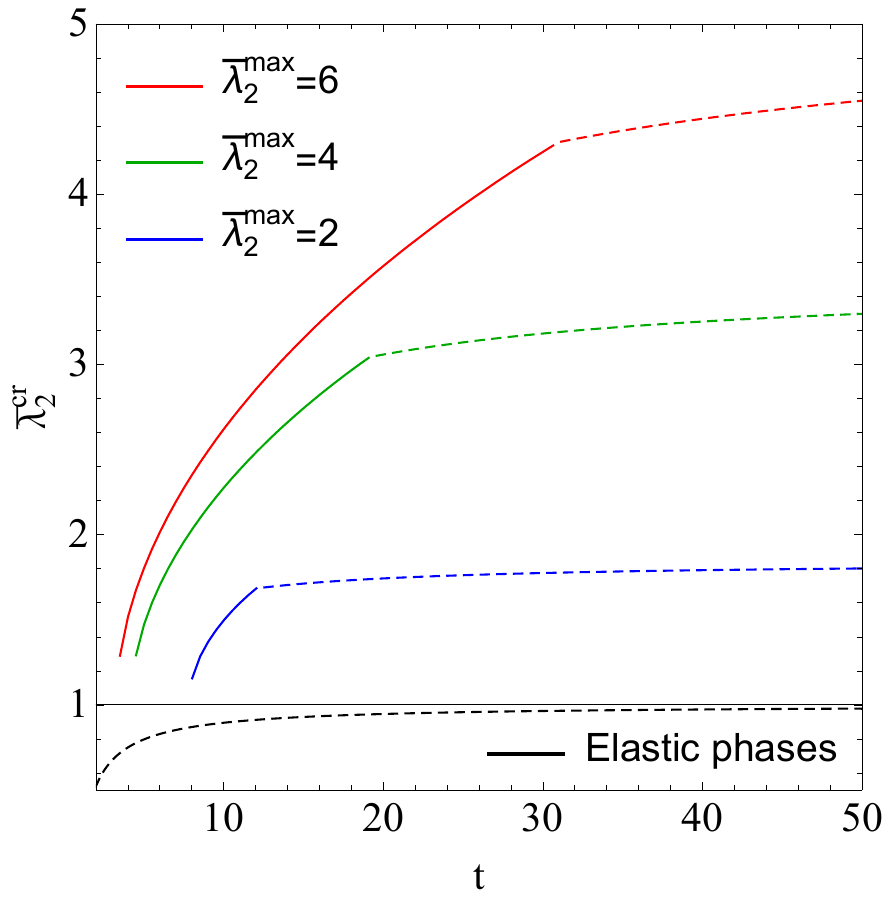} 
		& \includegraphics[width=3.25in]{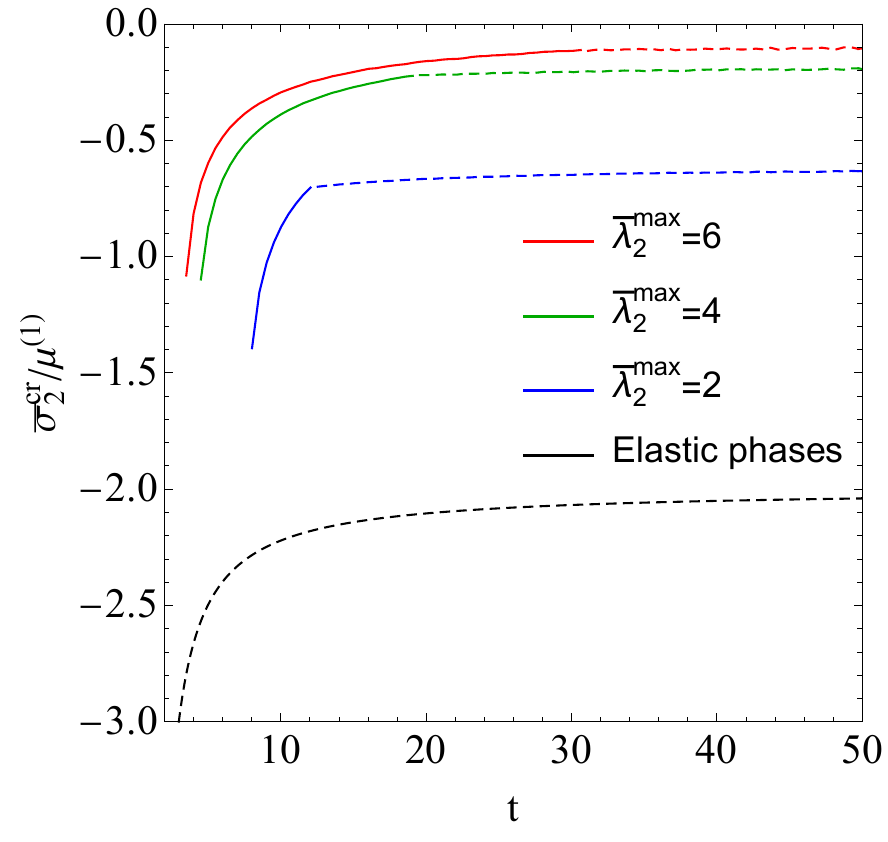} \\
		(a) & (b) \\
	\end{tabular}
	\caption{Results for the macroscopic LOE of elastoplastic laminates with phase volume fractions $c_0^{(1)}=c_0^{(2)}=0.5$ and local material properties $\kappa^{(1)} = 10 \mu^{(1)}$, $\Sigma^{(2)}_0=\mu^{(2)}/10$, and $h^{(2)} = \mu^{(2)}/5$, under aligned, plane-strain tension up to a maximum applied stretch $\ol^{max}_2$, followed by tensile unloading, and subsequent compressive loading. 
		(a) The macroscopic critical stretch $\ol^{cr}_{2}$ and (b) the corresponding normalized stress $\osig^{cr}_{2} / \mu^{(1)}$ are plotted as functions of the elastic phase contrast $t \equiv \mu^{(2)} / \mu^{(1)} = \kappa^{(2)} / \kappa^{(1)}$ for the cases $\ol^{max}_2 = 2$, $\ol^{max}_2 = 4$ and $\ol^{max}_2 = 6$. The associated critical curves for the corresponding laminates with purely elastic phases are also shown in this figure for comparison. Critical conditions corresponding to macroscopic LOE in the purely elastic and elasplastic regimes are shown with dashed and continuous lines, respectively.
	}
	\label{LOECurves_vs_t_L2max_6_4_2}
\end{figure}

Figure \ref{LOECurves_vs_t_L2max_6_4_2} shows plots of the macroscopic critical stretch $\ol^{cr}_{2}$ and normalized stress $\osig^{cr}_{2} / \mu^{(1)}$ as functions of the elastic contrast $t \equiv \mu^{(2)} / \mu^{(1)} = \kappa^{(2)} / \kappa^{(1)}$ for different values of the maximum applied stretch $\ol^{max}_2$. 
Corresponding plots (black dashed curves) are also shown in this figure for laminates with purely elastic, Neo-Hookean phases. 
It is observed that the variables $\ol^{cr}_{2}$ and $\osig^{cr}_{2}$ for the latter Neo-Hookean composites increase monotonically with increasing contrast $t$, suggesting that the softening effect induced on the critical mode by local elastic straining under decreasing loading conditions becomes progressively stronger with increasing $t$.
Note, in particular, that in the limit as $t \rightarrow \infty$, corresponding to laminates comprised of Neo-Hookean (phase 1) and approximately rigid (phase 2) layers, the loss of macroscopic ellipticity takes place at zero strain ($\ol^{cr}_{2} = 1$), but it requires the application of a non-zero stress ($\osig^{cr}_{2} < 0$), while in the other limit as $t \rightarrow 1$, corresponding to a homogeneous, Neo-Hookean solid, the material is strongly elliptic, as expected.
As opposed to their hyperelastic counterparts, the elastoplastic laminates of this figure do not lose strong ellipticity up to a certain value of $t$, depending on the maximum applied stretch $\ol^{max}_2$.
In particular, it is observed that the range of values of $t$ within which these composites remain strongly elliptic decreases with increasing $\ol^{max}_2$, which in turn reflects the greater softening induced on the critical mode upon tensile loading at higher values of $\ol^{max}_2$. 
Beyond these strongly elliptic regions, the elastoplastic composites of Fig. \ref{LOECurves_vs_t_L2max_6_4_2} lose macroscopic ellipticity continuously with increasing $t$, initially in the elastoplastic and subsequently in the purely elastic regime, and the corresponding critical conditions $\ol^{cr}_{2}$ and $\osig^{cr}_{2}$ increase monotonically with increasing $t$. 
The latter monotonic behavior of the critical curves for these elastoplastic composites is consistent with that of the corresponding curves for the Neo-Hookean laminates of this figure. As already mentioned, this behavior reflects the fact that the softening of the critical mode under decreasing local elastic straining becomes stronger at higher values of $t$.
It is also interesting to observe from the results of Fig. \ref{LOECurves_vs_t_L2max_6_4_2}(b) that the critical stress $\osig^{cr}_{2}$ tends to zero for increasing values of the maximum applied stretch $\ol^{max}_2$, but it remains compressive ($\osig^{cr}_{2} < 0$) for all values of $t$, which is consistent with the corresponding observation made earlier in the context of Fig. \ref{LOECurves_vs_L2max_h2_10_5_2_1}(b).

\section{Concluding remarks}
\label{CR}

%
Structural integrity of biological tissues, such as ligaments and tendons, is crucial for maintaining homeostasis and the load-bearing capacity of these materials.  Domain formation in these soft composites, when subjected to cyclic loading under natural operating conditions, leads often to severe and permanent changes in their microstructure and cellular microenvironment, and it is closely tied with the development of disease, such as tendinopathy. A detailed investigation of the onset and evolution of these domains requires taking into account the complexity of the underlying fibrous microstructure in these biological materials as well as the constitutive and geometric nonlinearities that are involved at the level of their constituents. 
In this work, we have been primarily concerned with the effect of fiber plasticity on the macroscopic response and domain formation in soft biological composites. 
For simplicity, we focused our considerations on the homogenized behavior of two-phase laminates, composed of purely elastic (matrix phase) and elastoplastic (fiber phase) layers. 
In addition, we restricted our analysis to plane-strain, non-monotonic loading conditions, aligned with the direction of the layers. 
In this context, we studied in detail the critical conditions for the onset of bifurcations of the long-wavelength type, as determined by the macroscopic loss of ellipticity condition, as well as the associated post-bifurcation behavior of these laminates. 
A simple expression has been obtained for the loss of macroscopic ellipticity in these composites.
It follows from this expression that, just like their hyperelastic counterparts, elastoplastic laminates lose macroscopic ellipticity whenever their incremental strength in shear perpendicular to the layers, as measured by the associated effective shear modulus $\tilde{\cal{L}}_{1212}$ of the laminate, vanishes for the first time.  
In addition, it has been shown that the post-bifurcation regime of these composites is characterized by the formation of twin lamellar domains at a mesoscopic length-scale, which is  much larger than the typical width of the original, microscopic layers, but still much smaller than the overall dimensions of the macroscopic specimen.
The two types of domains are layered along a direction parallel to the loading axis and realize at equal volume fractions ($50 \%$).
These findings are in agreement, at least qualitatively, with corresponding experimental evidence from studies on tendon (see, e.g., Fig. \ref{Tendon}). 
It has also been shown in this work by means of specific numerical calculations that domain formation acts as a stress-relaxation mechanism which allows macroscopic deformation to be accommodated primarily through microlayer rotation combined with intense longitudinal shear of the soft microlayers. 
As a result, the macroscopic response of the laminate right after the onset of bifurcation is much softer along the bifurcated path than along the corresponding principal path.
It should be emphasized, however, that no macroscopic unloading has been observed in the post-bifurcation regime of these elastoplastic composites, suggesting accordingly that the macroscopic, post-bifurcation behavior of these materials is stable.
These findings for the elastoplastic laminates under consideration here are in qualitative agreement with the corresponding results of \cite{FPC18} for laminates with purely Neo-Hookean phases.

The influence of local plastic deformations on the onset of bifurcations of the of long-wavelength type in the lamellar composites of interest has been illustrated by means of detailed numerical calculations.
It has been found that fiber plasticity has a dominant softening effect on the incremental shear modulus $\tilde{\cal{L}}_{1212}$ of the laminate under increasing loading and a hardening effect under decreasing loading, while local elasticity has the opposite effects. 
In other words, the effects induced by local elasticity and fiber plasticity on the effective modulus $\tilde{\cal{L}}_{1212}$ compete with each other at every stage of the deformation where both mechanisms are active, and whether or not a specific laminate will lose its macroscopic ellipticity along a given loading path depends crucially on the relative strengths of these mechanisms.
The results that have been generated for the purpose of this work show consistently that the modulus $\tilde{\cal{L}}_{1212}$ does never vanish during tensile loading or unloading. 
For these reasons, we conclude that the elastoplastic laminates under consideration may lose macroscopic ellipticity only during the compressive loading stage. 
It has also been found that elastoplastic laminates may lose macroscopic ellipticity under extensive stretches ($\ol^{cr}_2 > 1$) along the direction of the layers, which is in sharp contrast with the fact that corresponding hyperelastic laminates may lose macroscopic ellipticity only under contractile stretches ($\ol^{cr}_2 < 1$) (or at $\ol^{cr}_2 = 1$ for the special case when one of the phases is rigid).
It has also been shown in this work that the effects induced by fiber plasticity on $\tilde{\cal{L}}_{1212}$ become stronger for lower values of the initial yield stress $\Sigma^{(2)}_0$ and/or the instantaneous hardening rate $h^{(2)}$ of the elastoplastic phase, which is consistent with the fact that decreasing $\Sigma^{(2)}_0$ and/or $h^{(2)}$ leads to increasing local plastic yielding and decreasing local elastic straining.
As a result, laminates with large and moderate values of $\Sigma^{(2)}_0$ and $h^{(2)}$ lose macroscopic ellipticity under incrementally elastic and elastoplastic deformations, respectively. 
On the other hand, laminates with sufficiently small values of $\Sigma^{(2)}_0$ and $h^{(2)}$ remain strongly elliptic throughout the loading path,
provided that the strain hardening of the fiber phase is strong enough to maintain the strongly elliptic behavior of this phase. Composites with nearly ideally plastic fibers ($h^{(2)} \approx 0$) have been found to lose strong ellipticity locally, in this elastoplastic phase, shortly after the onset of plastic yielding during tension. The potential connection of this local loss of ellipticity with the micro-necking of collagen fibers that has been observed in tendons (Fig. \ref{Tendon}(c)) is currently under investigation and will be discussed elsewhere.
Another interesting conclusion that can be drawn from the present study is that increasing the maximum applied stretch $\ol^{max}_2$ and, thus, the total amount of local plastic deformations during the tensile loading stage, makes the laminate more prone to lose macroscopic ellipticity, in the sense that,
whenever macroscopic LOE does take place, the corresponding macroscopic critical stretch $\ol^{cr}_2$ and stress $\osig^{cr}_2$ along the layers increase monotonically with increasing $\ol^{max}_2$. 
In this connection, it should be emphasized once again that the elastoplastic laminates under consideration here have not been found to lose macroscopic ellipticity under macroscopically tensile stresses along the direction of the layers. 
By contrast, domain formation in actual biological materials, like tendons, is known to occur under tensile loading/unloading cycles along the fiber direction. 
Thus, it may be inferred from the results of this work that fiber plasticity may have a significant effect in increasing the applied critical stress along the fiber direction towards zero, but this effect is most likely not enough to fully account for domain formation in tendons.
This point is further investigated in the context of the imperfection analysis that is carried out in a forthcoming (Part II).
Additional micro-mechanisms which may also play an important role in this phenomenon, but which have not been accounted for in the context of the present analysis, include the damage that might take place in the fiber phase of the actual composites.
A detailed investigation along these lines is currently underway and will be reported elsewhere.

\section{Acknowledgments}

M.A. and N.B. thank Professor Pedro Ponte Casta\~neda for pointing out the connection of some of the early findings of this work with his research concerning domain formation in hyperelastic composites.
M.A. also thanks Professor Nikolaos Aravas for fruitful discussions and recommendations. 
F.F.F. and N.B. acknowledge the support by the National Science Foundation under grant
no. CMMI-2038057.

\section*{Appendix: Finite elastic and plastic deformations}
\label{FDef}

In this appendix, following \cite{PAB1993}, we define constitutive relations for homogeneous solids undergoing finite elastic and plastic deformations. 
These relations are used in the main body of the article to characterize the local material behavior in the constituent phases of the composite materials of interest.

\subsection*{Kinematics}
\label{Kinematics}

Consider an arbitrary deformation $\bfx = \bfx (\bfX,t)$ of a homogeneous body from 
a fixed reference configuration $\Omega_0$ at time $t_0$ to its current configuration $\Omega$ at time $t$, where the position vectors of a typical material point are denoted by $\bfX$ and $\bfx$, respectively. Let $\bfv = \dot{\bfx} (\bfX,t)$ denote the corresponding velocity field. The mapping $\bfx = \bfx (\bfX,t)$ is assumed to be one-to-one and differentiable, so that the deformation gradient tensor $\bfF$ and the velocity gradient tensor $\bfL = \dot{\bfF} \bfF^{-1}$, with components $F_{ij} = \p x_i / \p X_j$ and $L_{ij} = \p v_i / \p x_j = \dot{F}_{ik} F^{-1}_{kj}$, respectively, are well defined at each point. In addition, the deformation is required to satisfy the impenetrability condition $J \equiv \det \bfF > 0$.

Following Lee (1969), it is further assumed that, at any given point $\bfX$ in $\Omega_0$, the tensor $\bfF$ may be decomposed into a plastic part $\bfF^p$ and an elastic part $\bfF^e$ as follows
\begin{equation}
	\bfF = \bfF^e \bfF^p,
	\label{FeFp}
\end{equation}
where both mappings $\bfF^p$ and $\bfF^e$ are taken to be invertible and to have positive determinants, i.e., $J^p \equiv \det \bfF^p > 0$ and $J^e \equiv \det \bfF^e > 0$.
Note that, the decomposition (\ref{FeFp}) introduces a local intermediate configuration\footnote{Equation (\ref{FeFp}) defines the intermediate configuration only up to a rigid body rotation. This ambiguity does not affect the formulation of the constitutive equations for the isotropic materials under consideration in this work. Detailed discussions concerning the proper choice of this configuration for the case of anisotropic materials can be found in the work by \cite{PAB1993} and in references therein.} which may be regarded at the same time as the deformed state for the plastic part and as the reference state for the elastic part of the deformation. 
As detailed further below, the constitutive equations and symmetries for both the elastic and the plastic part of the material behavior may be conveniently stated in the intermediate configuration.

Elastic and plastic strain measures may thus be defined in terms of $\bfF^e$ and $\bfF^p$, respectively. In this work, we will only make use of the Green elastic strain tensor
\begin{equation}
	\bfE^e = \frac{1}{2} \left( \bfC^e - \bfdel \right),
	\label{Ee}
\end{equation}
where $\bfC^e =  \bfF^{e T} \bfF^e$ is the elastic right Cauchy-Green deformation tensor and $\bfdel$ is the second-order identity tensor. 

Taking into account (\ref{FeFp}), the velocity gradient tensor $\bfL = \dot{\bfF} \bfF^{-1}$ takes the form
\begin{equation}
	\dot{\bfF}^e \bfF^{e-1} + \bfF^e \bfL^i  \bfF^{e-1}, \quad \quad 
	\bfL^i = \dot{\bfF}^p \bfF^{p-1},
	\label{L2} 
\end{equation}
where $\bfL^i$ is the velocity gradient in the intermediate configuration. The tensors $\bfL$ and $\bfL^i$ may be decomposed into the deformation rate tensors
\begin{equation}
	\bfD = \bfL_s \equiv \bfD^e + \bfD^p, \quad\quad 
	\bfD^e = \left( \dot{\bfF}^e \bfF^{e-1} \right)_s, \quad \quad
	\bfD^p = \left( \bfF^e \bfL^i \bfF^{e-1} \right)_s, 
	\quad \quad
	\bfD^i = \bfL^i_s,  
	\label{D}
\end{equation}
and the spin tensors 
\begin{equation}
	\bfW = \bfL_a \equiv \bfW^e + \bfW^p, \quad \quad 
	\bfW^e = \left( \dot{\bfF}^e \bfF^{e-1} \right)_a, \quad \quad
	\bfW^p = \left( \bfF^e \bfL^i \bfF^{e-1} \right)_a, 
	\quad \quad
	\bfW^i = \bfL^i_a,  
	\label{W}
\end{equation}
where the subscripts ``s'' and ``a'' indicate respectively the symmetric and anti-symmetric part of the second-order tensor to which they are attached.

\subsection*{Stress power}
\label{SP}

It follows from the above definitions that the stress power $\mathcal{P}$ per unit volume of intermediate configuration may be decomposed accordingly into an elastic part $\mathcal{P}^e$ and a plastic part $\mathcal{P}^p$ as follows
\begin{equation}
	\mathcal{P} = \bftau \cdot \bfD \equiv \mathcal{P}^e + \mathcal{P}^p, \quad\quad  \mathcal{P}^e = \bfS \cdot \dot{\bfE}^e, \quad\quad  \mathcal{P}^p = \bfSig \cdot \bfL^i,
	\label{StressPower}
\end{equation}
where we have introduced the stress measures 
\begin{equation}
	\bftau = J^e \bfsig, \quad\quad
	\bfS = \bfF^{e-1} \bftau \bfF^{e -T}, \quad\quad \bfSig = \bfF^{e T} \bftau \bfF^{e -T},
	\label{StressTensors}
\end{equation}
with $\bfsig$ denoting the Cauchy stress tensor. The tensors $\bftau$, $\bfS$, and $\bfSig$ in (\ref{StressTensors}) are commonly referred to as the (elastic) Kirchhoff, the second Piola-Kirchhoff, and the Mandel stress, respectively. The Cauchy stress $\bfsig$ is assumed to be symmetric, so that the tensors $\bftau$ and $\bfS$ are also symmetric, but $\bfSig$ is not symmetric, in general.

In this work, we focus our considerations on materials that are elastically isotropic with respect to the intermediate configuration. For this special class of materials, the Mandel stress tensor $\bfSig$ is also symmetric\footnote{This fact follows easily from $(\ref{StressTensors})_3$, which for this special case reduces to $\bfSig = \bfR_e^T \bftau \bfR_e = \bfSig^T$, where $\bfR_e$ is the rotation tensor from the polar decomposition of $\bfF^e$} and, therefore, the associated plastic stress power $(\ref{StressPower})_3$ reduces to
\begin{equation}
	\mathcal{P}^p = \bfSig \cdot \bfD^i.
	\label{PpIso}
\end{equation}
In addition, we restrict our attention on plastically incompressible materials, for which, taking into account the constraint $J^p = 1$, it follows that $J = J^e$ and that the total stress power $(\ref{StressPower})_1$ may also be written in the form
\begin{equation}
	\mathcal{P} = \bfP \cdot \dot{\bfF}, 
	\label{StressPowerRef}
\end{equation}
where 
\begin{equation}
	\bfP = \bftau \bfF^{-T} = J \bfsig \bfF^{-T}, 
	\label{FirstPK}
\end{equation}
is the first Piola-Kirchhoff stress tensor.

In the sequel, assuming that $(\ref{PpIso})$ and $(\ref{StressPowerRef})$ hold, we define the elastic and plastic part of the material behavior in the intermediate configuration. In particular, we first establish appropriate relations between the conjugate variables $(\bfS, \bfE^e)$ and $(\bfSig, \bfD^i)$, and then we combine these relations to obtain the  corresponding rate form of the elastoplastic equation for the conjugate pair $(\bfP, \bfF)$.

\subsection*{Elasticity}
\label{Elasticity}

The elastic properties of the material are assumed to be given in terms of an isotropic strain-energy density function such that $\mathcal{P}^e = \dot{\Psi}$ and, therefore,
\begin{equation}
	\bfS = \frac{\p \Psi}{\p \bfE^e}.
	\label{Sx}
\end{equation}
The above constitutive equation may also be written in the following rate 
form 
\begin{equation}
	\dot{\bfS} = \bsC_e \dot{\bfE}^e, \quad\quad \bsC_e =\frac{\p^2 \Psi}{\p \bfE^e \p \bfE^e}.
	\label{Sxdot}
\end{equation}
Making use of $(\ref{StressTensors})_2$ and $(\ref{Sxdot})$, it is straightforward to show that 
\begin{equation}
	\bftaut = \bsL_e \bfD^e, \quad\quad 
	\left( \bsL_e \right)_{ijkl} = F^e_{ip} F^e_{jq} F^e_{kr} F^e_{ls} \left( \bsC_e \right)_{pqrs} +
	\frac{1}{2} \left( \tau_{ik} \delta_{jl} + \tau_{il} \delta_{jk} + \delta_{ik} \tau_{jl} + \delta_{il} \tau_{jk}  \right),
	\label{taut}
\end{equation}
where we have introduced the notation $\bftaut = \dot{\bftau} + \bftau \bfW^e - \bfW^e \bftau$. Note that, the fourth-order modulus tensors $\bsL_e$ and $\bsC_e$ in (\ref{taut}) and (\ref{Sxdot}) possess both the major and the minor symmetries.

\subsection*{Plasticity}
\label{Plasticity}

The plastic part of the material behavior, which is also assumed to be isotropic with respect to the intermediate configuration, is defined in terms of the following von Mises yield criterion with isotropic strain hardening
\begin{equation}
	\Phi (\bfSig, \eps^p) = \Sigma_{eq} - \Sigma_{y} (\eps^p) = 0, \quad\quad 
	\Sigma_{eq} = \sqrt{\frac{3}{2} \bfSig^{d} \cdot \bfSig^{d}},
	\label{YC}
\end{equation}
where $\bfSig^{d}$ is the deviatoric part of the Mandel stress tensor $\bfSig$ and $\Sigma_{y}$ is the associated yield stress, which is given as a function of the equivalent plastic strain $\eps^p$, to be determined further below.

The purely plastic stretching tensor $\bfD^i$ in the intermediate configuration is determined in terms of its conjugate stress tensor $\bfSig$ (see equation (\ref{PpIso})) by the associative flow rule  
\begin{equation}
	\bfD^i = \dot{\Lambda} \bfM^i, \quad\quad \bfM^i = \frac{\p \Phi}{\p \bfSig} = \frac{3}{2 \Sigma_{eq}} \bfSig^{d},
	\label{Di}
\end{equation}
where the plastic multiplier $\dot{\Lambda} \ge 0$ is obtained from the consistency condition, as discussed further below. 
Taking into account the assumed isotropic plastic behavior, it follows that (see, e.g., \cite{PAB1993})
\begin{equation}
	\bfW^i = \bfzero.
	\label{Wi}
\end{equation}
Hence, the deformation gradient in the intermediate configuration $\bfL^i = \dot{\bfF}^p \bfF^{p-1} = \bfD^i$ is symmetric. 

Given $(\ref{Di})$ and $(\ref{Wi})$, the plastic stretching tensor $(\ref{D})_3$ in the current configuration takes the form
\begin{equation}
	\bfD^p = \dot{\Lambda} \bfM^p, \quad\quad \bfM^p = \left( \bfF^e \bfM^i \bfF^{e-1} \right)_s.
	\label{Dp}
\end{equation}
In addition, it follows from the representation theorems for isotropic functions and the elastic isotropy of the material that $\bfW^p = \bfzero$. Therefore, $\bfW = \bfW^e$ and the tensor $\bfF^e \bfL^i \bfF^{e-1} = \bfF^e \bfD^i \bfF^{e-1} = \dot{\Lambda} \bfF^e \bfM^i \bfF^{e-1}$ is symmetric.

The equivalent plastic strain $\eps^p$ is taken to be plastic-work conjugate to the equivalent stress $\Sigma_{eq}$, so that
\begin{equation}
	\mathcal{P}^p = \bfSig \cdot \bfD^i = \Sigma_{eq} \dot{\eps}^p.
	\label{PpIso2}
\end{equation}
It follows that 
\begin{equation}
	\dot{\eps}^p = \dot{\Lambda},
	\label{epdot}
\end{equation}
which constitutes the evolution equation for $\eps^p$.

Given the yield condition $(\ref{YC})$ and the above evolution equation $(\ref{epdot})$, the consistence condition $\dot{\Phi} = 0$ takes the form
\begin{equation}
	\bfM^i \cdot \dot{\bfSig} - \dot{\Lambda} h = 0, \quad\quad h = \frac{d \Sigma_y}{d \eps^p}.
	\label{CC}
\end{equation}
Taking the time derivative of $(\ref{StressTensors})_3$ with respect to time, and making use of the result $(\ref{taut})$, it follows that 
\begin{equation}
	\dot{\bfSig} = \bfF^{e T} \left[ \left( \bsL_e + \bsT \right) \bfD^e \right] \bfF^{e -T},
	\label{Sigmadot}
\end{equation}
where
\begin{equation}
	\scT_{ijkl} = \frac{1}{2} \left( -\tau_{ik} \delta_{jl} - \tau_{il} \delta_{jk} + \delta_{ik} \tau_{jl} + \delta_{il} \tau_{jk} \right).
	\label{T}
\end{equation}
Note that the tensor $\bsT$ is anti-symmetric with respect to the first and symmetric with respect to the second pair of indexes.
Substituting expression $(\ref{Sigmadot})$ in the consistency condition $(\ref{CC})$, and making use of the fact that $\bfD^e = \bfD - \bfD^p$, with $\bfD^p$ given by $(\ref{Dp})$, 
it can be shown that 
\begin{equation}
	\dot{\Lambda} = \bfr \cdot \bfD,
	\label{ldot}
\end{equation}
where
\begin{equation}
	\bfr = \frac{1}{L} \left( \bfF^e  \bfM^i \bfF^{e-1} \right) \left( \bsL_e + \bsT \right), \quad\quad  
	L= \left( \bfF^e  \bfM^i \bfF^{e-1} \right) \cdot \left( \bsL_e + \bsT \right) \bfM^p + h.
	\label{rL}
\end{equation}
Given that the tensor $\bfF^e  \bfM^i \bfF^{e-1} = \bfM^p$ is symmetric and $\bsT$ is antisymmetric with respect to the first pair of indexes, it follows that $\left( \bfF^e  \bfM^i \bfF^{e-1} \right) \bsT = \bfzero$. Thus, equation $(\ref{ldot})$, with $(\ref{rL})$, simplifies to 
\begin{equation}
	\dot{\Lambda} = \frac{1}{L} \bfM^p \cdot \bsL_e \bfD, \quad\quad  
	L= \bfM^p \cdot \bsL_e \bfM^p + h.
	\label{ldot2}
\end{equation}

\subsection*{Incremental constitutive equations}
\label{IncrementalEqs}

Taking into account the fact that $\bfD^e = \bfD - \bfD^p$, as well as equation $(\ref{Dp})$ for $\bfD^p$, with $\dot{\Lambda}$ given by $(\ref{ldot2})$, the rate equation $(\ref{taut})$ may be recast in the form
\begin{equation}
	\bftauj = \bsL \bfD, \quad\quad 
	\bsL = \bsL_{e} -  \frac{u(\dot{\Lambda})}{L} \bsL_e \bfM^p \otimes \bfM^p \bsL_e
	\label{tauj}
\end{equation}
where $\bftauj = \dot{\bftau}- \bfW \bftau + \bftau \bfW $ is the Jaumann derivative of $\bftau$ and $u(\dot{\Lambda})$ is the unit step function; recall that $u(\dot{\Lambda})=1$ if $\dot{\Lambda} > 0$ and $u(\dot{\Lambda})=0$ otherwise. Note that, in the context of $(\ref{tauj})$, use has also been made of the fact that $\bfW=\bfW^e$ for the isotropic materials under consideration, which in turn implies that $\bftaut = \bftauj$. 
Note also that, the modulus tensor $\bsL$ in  $(\ref{tauj})_2$ has both the major and the minor symmetries.

Differentiating equation $(\ref{StressPowerRef})_2$ with respect to time and making use of the constitutive relation $(\ref{tauj})$, it is straightforward to show that
\begin{equation}
	\dot{\bfP} = \bmL \, \dot{\bfF}, \quad\quad 
	\mathcal{L}_{ijkl} = \left[\mathscr{L}_{ipkq} +  
	\frac{1}{2} \left( \delta_{ik} \tau_{pq} - \delta_{iq} \tau_{kp}  -
	\delta_{pq} \tau_{ik}  - \delta_{kp} \tau_{iq} \right) \right] F^{-1}_{jp} F^{-1}_{lq}.
	\label{Sdot}
\end{equation}
It is emphasized that the elastoplastic modulus tensor $\bmL$ in $(\ref{Sdot})$ depends on the current state of the material, as defined by the variables $\bfF^p$, $\bfF^e$, and $\eps^p$, as well as on the loading increment (through its dependence on $\dot{\Lambda}$). Note that, the modulus tensor $\bmL$ has the major symmetry, but it does not have the minor symmetries. For purely elastic increments, expression $(\ref{Sdot})_2$ takes the form 
\begin{equation}
	\mathcal{L}_{ijkl} = \left[ F^e_{ip} F^e_{kr} (\bsC_{e})_{pqrs} + \delta_{ik} S_{qs} \right] (\bfF^{p-1})_{jq} (\bfF^{p-1})_{ls},
	\label{LepEl}
\end{equation}
where we recall that the tensors $\bfS$ and $\bsC_{e}$ are respectively given by $(\ref{Sx})$ and $(\ref{Sxdot})_2$ in terms of the Green elastic strain tensor $\bfE^e$. For the special case of purely elastic deformations, i.e., for $\bfF = \bfF^e$, expression $(\ref{LepEl})$ reduces to
\begin{equation}
	\mathcal{L}_{ijkl} = F^e_{ip}  F^e_{kr}  (\bsC_{e})_{pjrl} + \delta_{ik} S_{jl} = \frac{\p \psi}{\p \bfF^e \p \bfF^e},
	\label{LeEl}
\end{equation}
as it should.

\subsection*{Numerical integration of the incremental constitutive equations}
\label{NI}

For completeness, an algorithm for the numerical integration of the elastoplastic constitutive equations of this Appendix is presented next.

Let $\Delta t = t_{n+1} - t_n$, with $n=0, 1, 2,  ...$, be the generic time increment
during an arbitrary homogeneous deformation history, prescribed by the deformation gradient tensor $\bfF = \bfF (t)$. Given the values of the variables $\bfF^e$, $\bfF^p$ and $\eps^p$ at time $t_n$, as well as the value of $\bfF$ at time $t_{n+1}$, the prime goal is to find the values of $\bfF^e$, $\bfF^p$ and $\eps^p$ at $t_{n+1}$. To this end, note first that, since $\bfD^i = \dot{\bfF}^p \bfF^{p-1}$, the normality rule (\ref{Di}) may be rewritten in the form 
\begin{equation}
	\dot{\bfF}^p = \dot{\Lambda} \bfM^i \bfF^p
	\label{Fpdot}
\end{equation}
Assuming that the direction of the plastic flow $\bfM^i$ is constant and equal to $\bfM_{i} (t_n)$ within the time increment $\Delta t$, and integrating the ODE (\ref{Fpdot}) over $\Delta t$, we obtain the result
\begin{equation}
	\bfF^{p-1}(t_{n+1}) =
	\bfF^{p-1}(t_{n}) \left[\bfdel -  \Delta \Lambda \bfM_{i}(t_{n}) + \frac{1}{2} \Delta \Lambda^2 \bfM_{i}^2(t_{n}) +  O\left( \Delta \Lambda^3 \bfM_{i}^3(t_{n}) \right)  \right]
	\label{Fpnp1}
\end{equation}
which, truncated to second-order in $\Delta \Lambda$, may be used to compute $\bfF_{p}(t_{n+1})$ in terms of $\Delta \Lambda$. Given this result, $\bfF^e(t_{n+1})$ may also be obtained in terms  of $\Delta \Lambda$ from the expression
\begin{equation}
	\bfF^e(t_{n+1}) = \bfF(t_{n+1}) \bfF^{p-1}(t_{n+1}).
	\label{Fenp1}
\end{equation}
In addition, integrating the evolution equation (\ref{epdot}), we find that 
\begin{equation}
	\eps^p(t_{n+1}) = \eps^p(t_{n}) + \Delta \Lambda.
	\label{epnp1}
\end{equation}
Hence, the unknown variables $\bfF^p(t_{n+1})$, $\bfF^e(t_{n+1})$, and $\eps^p(t_{n+1})$ may be readily determined from (\ref{Fpnp1}), (\ref{Fenp1}), and (\ref{epnp1}), respectively, in terms of $\Delta \Lambda$.

In order to find the principal unknown $\Delta \Lambda$, we assume first that the deformations within the increment $\Delta t$ are purely elastic, so that $\Delta \Lambda = 0$, and compute the variables
\begin{equation}
	\bfC^e(t_{n+1}) = \bfF^{e T}(t_{n+1}) \bfF^e(t_{n+1}), \quad\quad
	\bfE^e(t_{n+1}) = \frac{1}{2} \left[ \bfC^e(t_{n+1}) - \bfdel \right],
\end{equation}
\begin{equation}
	\bfS(t_{n+1}) = \frac{\p \Psi}{\p \bfE^e} \left[ \bfE^e( t_{n+1}) \right], \quad\quad
	\bfSig( t_{n+1}) = \bfC^e(t_{n+1}) \bfS(t_{n+1}), \quad\quad
	\Sigma_{eq} \left[\bfSig( t_{n+1})\right] = \sqrt{\frac{3}{2} \bfSig^{d}( t_{n+1}) \cdot \bfSig^{d}(t_{n+1})},
	\label{Snp1}
\end{equation}
\begin{equation}
	\Phi \left[\bfSig( t_{n+1}), \eps^p( t_{n})\right] = \Sigma_{eq} \left[\bfSig( t_{n+1})\right] - \Sigma_{y} \left[\eps^p( t_{n})\right].
\end{equation}
If $\Phi \left[\bfSig( t_{n+1}), \eps^p( t_{n})\right] < 0$, the deformations within $\Delta t$ are indeed purely elastic and the unknown variables at the end of the increment are $\bfF^p(t_{n+1}) = \bfF^p(t_{n})$, $\bfF^e(t_{n+1}) = \bfF(t_{n+1}) \bfF^{p-1}(t_{n})$, and $\eps^p(t_{n+1}) = \eps^p(t_{n})$. If, on the other hand, $\Phi \left[\bfSig( t_{n+1}), \eps^p( t_{n})\right] \ge 0$ the principal unknown $\Delta \Lambda$ is obtained by solving the yield condition at the end of the increment: 
\begin{equation}
	\Phi \left[\bfSig( t_{n+1}), \eps^p( t_{n+1})\right] = \Sigma_{eq} \left[\bfSig( t_{n+1})\right] - \Sigma_{y} \left[\eps^p( t_{n+1})\right] = 0,
	\label{YCnp1}
\end{equation}
where we recall that the variables $\bfSig( t_{n+1})$ and $\eps^p( t_{n+1})$ are determined in term of $\Delta \Lambda$ by means of the expressions (\ref{Fpnp1})-(\ref{Snp1}). Equation (\ref{YCnp1}) is nonlinear and must be solved numerically for $\Delta \Lambda$, e.g., by means of Newton's method.
Given $\bfF( t_{n+1})$, $\bfF^e( t_{n+1})$, and $\bfSig( t_{n+1})$, the stress tensors $\bftau$, $\bfsig$, and $\bfP$ may be determined at time $t_{n+1}$ by making use of the equations $(\ref{StressTensors})_3$, $(\ref{StressTensors})_1$, and $(\ref{FirstPK})$ respectively.

\bibliographystyle{unsrt}  
\bibliography{references}  






\end{document}